\documentclass{aa}
\usepackage{txfonts}
\usepackage{float}
\usepackage{graphicx} 
\usepackage{caption}
\usepackage{subcaption}
\usepackage{enumitem}

\begin{document} 

\authorrunning{E. Samara et al.,}
\titlerunning{Implementing the MULTI-VP coronal model in EUHFORIA}

\title{Implementing the MULTI-VP coronal model in EUHFORIA:\\ test case results and comparisons with the WSA coronal model}
   
   \author{E. Samara\inst{1,2}, R. F. Pinto\inst{3,4}, J. Magdaleni\'{c}\inst{1,2}, N. Wijsen\inst{2}, V. Jer\v{c}i\'c\inst{2}, C. Scolini\inst{2,1}, I. C. Jebaraj\inst{1,2}, \\  L. Rodriguez\inst{2}, S. Poedts\inst{2,5}}
   
   \institute{1.~SIDC, Royal Observatory of Belgium, Brussels, Belgium\\
             2.~CmPA, KU Leuven, Leuven, Belgium \\
             3.~IRAP, Université de Toulouse, CNRS, UPS, CNES, Toulouse, France\\
             4.~LDE3, CEA Saclay, Université Paris-Saclay, Gif-sur-Yvette, France\\
             5.~Institute of Physics, University of Maria Curie-Sk{\l}odowska, Lublin, Poland}

\date{Accepted for publication in $Astronomy$ $\text{\&}$ $Astrophysics$ on February 5, 2021}

    \abstract 
   {In this study, we focus on improving EUHFORIA (European Heliospheric Forecasting Information Asset), a recently developed 3D MHD space weather prediction tool. EUHFORIA consists of two parts, covering two spatial domains; the solar corona and the inner heliosphere. For the first part, the semi-empirical Wang-Sheeley-Arge (WSA) model is used by default, which employs the Potential Field Source Surface (PFSS) and Schatten Current Sheet (SCS) models to provide the necessary solar wind plasma and magnetic conditions above the solar surface, at $0.1$~AU, that serve as boundary conditions for the inner heliospheric part. Herein, we present the first results of the implementation of an alternative coronal model in EUHFORIA, the so-called MULTI-VP model.}
   {After we replace the default EUHFORIA coronal set-up with the MULTI-VP model, we compare their outputs both at $0.1$~AU and $1$~AU, for test cases involving high speed wind streams (HSSs). We select two distinct cases where the standard EUHFORIA set-up failed to reproduce the HSS plasma/magnetic signatures at Earth, in order to test the performance of MULTI-VP coupled with EUHFORIA-heliosphere.}
   {To understand the quality of modeling with MULTI-VP in comparison with the default coronal model in EUHFORIA, we considered one HSS case during a period of low solar activity and another one during a period of high solar activity. Moreover, the modeling of the two HSSs was performed by employing magnetograms from different providers; one from the Global Oscillation Network Group (GONG) and the second from the Wilcox Space Observatory (WSO). This way, we were able to distinguish differences arising not only because of the different models but also because of different magnetograms.}
   {The results indicate that when employing a GONG magnetogram, the combination MULTI-VP+EUHFORIA-heliosphere reproduces the majority of HSS plasma and magnetic signatures measured at L1. On the contrary, the standard WSA+EUHFORIA-heliosphere combination does not capture the arrival of the HSS cases at L1. When employing WSO magnetograms, MULTI-VP+EUHFORIA-heliosphere reproduces the HSS that occurred during the period of high solar activity while it is ambiguous if it models the HSS during the period of low solar activity. For the same magnetogram and periods of time, WSA+EUHFORIA-heliosphere is not able to capture the HSSs of interest.}
   {The results show that the accuracy of the simulation output at Earth depends on the choice of both the coronal model and the input magnetogram. Nevertheless, a more extensive statistical analysis is necessary to determine how precisely these choices affect the quality of the solar wind predictions.}

               
\maketitle

\section{Introduction} \label{Sec:Introduction}

Over the last decades, the dependence of our society on technological assets and systems has been rapidly increasing. The possible impact of severe space weather conditions on technology raises concerns about the economic risks in case of extreme space weather events \citep{schrijver15, Knipp2018}. To protect the technological infrastructure on which society is ever-more dependent, reliable and timely space weather forecasting is required to enable mitigation scenarios. As the majority of the presently available models and tools do not consistently provide forecasts with sufficient accuracy, it is necessary to develop next-generation, more reliable space weather prediction tools. These tools may help in mitigating the effects of space weather on ground-based and space technological systems, including current and future space missions. EUHFORIA ("European Heliospheric Forecasting Information Asset", \citet{pomoell18}) is a recently developed, physics-based, 3D magnetohydrodynamics (MHD) forecasting-targeted model that aims to provide accurate solar wind and coronal mass ejection (CME) predictions at Earth and at any other point of interest within the inner heliosphere. 

Over the years, two distinct types of solar wind have been identified having different properties and sources, i.e.\ fast and slow solar wind \citep{schwenn06, cranmer17, McComas98}. Fast solar wind, often referred to as high speed streams (HSSs), can cause geomagnetic storms and can drive intense space weather conditions at Earth \citep{Richardson01, vrsnak07, hofmeister18}. To identify the arrival of fast solar wind streams at Earth, we look for a simultaneous increase of the solar wind speed, magnetic field and temperature, preceded by a density increase (due to the compression of the upcoming fast solar wind with the preceding slow solar wind). Additionally, the longitudinal angle of the interplanetary magnetic field vector, (the so-called IMF $\phi$-angle) should indicate a predominant direction (depending on whether the field is pointing towards or away from the Sun). This is also the criteria we followed to identify the HSSs in this work, for velocities exceeding 400 km/s.

CMEs are magnetic structures propagating within the solar wind. They are considered to be the major drivers of space weather at Earth and are responsible for the largest geomagnetic storms \citep{Webb2000, Hudson06, Gopalswamy05, richardson10, Kilpua2017, Wu2019, scolini19, Verbeke2019}. As both the solar wind and the CMEs are magnetised, their interaction strongly affects the evolution of the CMEs, such as their deviation, deformation and erosion \citep{Odstrcil04, Lugaz12, scolini2020}. From the modeling point of view, we identify many cases with EUHFORIA in which the simulation of CMEs propagating through a non-realistic solar wind leads to unreliable space weather predictions. Therefore, realistic modeling of solar wind is imperative not only for forecasting medium/large-scale geomagnetic storms caused by fast streams (predominantly during periods of low solar activity, see \citet[][]{Richardson2000})\footnote{For large geomagnetic storms, $Kp$ index ranges between 7$_{-}$ $\leq$ $Kp_{max}$ $\leq$ 7$_{+}$ and $Kp$ $\geq$ 6$_{-}$ for at least three 3-hour intervals in a 24-hour period. For medium-scale geomagnetic storms, $Kp_{max}$ $\geq$ 6$_{-}$.} but also for CMEs.

First results of solar wind modeling with EUHFORIA show that the velocities of both the slow and fast solar wind are frequently underestimated \citep{Hinterreiter19}. Improving the background solar wind modeled by EUHFORIA is a multi-dimensional problem, involving a number of different factors. For example, magnetograms from different sources, different coronal models, and different initial input parameters to the model (such as the initial density of the solar wind or the source surface height of the PFSS model) can lead to diverse results. 

In the present paper, we will focus on the implementation of an alternative coronal model in EUHFORIA, the MULTI-VP model \citep{Pinto17}. The coupling and testing of alternative coronal models, other than the default one, is essential towards achieving an optimal set-up for both background solar wind parameters and CME evolution. After we describe the heliospheric boundary condition requirements, we will present and discuss the modeled output at $0.1$~AU and $1$~AU. We note that the current analysis is meant to provide only a qualitative overview of the performance of different magnetograms and coronal models used in EUHFORIA for two selected HSSs at the Lagrangian point 1 (L1). A quantitative comparison will be the subject of a future work that will specifically focus on evaluating the differences of the simulation output from different coronal models and magnetograms, using appropriate metrics.

The paper is structured as follows. In Section 2, we describe the default set-up of EUHFORIA and MULTI-VP. In section 3, we discuss the coupling of MULTI-VP to the heliospheric part of EUHFORIA. Section 4 contains a presentation of results for plasma and magnetic parameters at the inner boundary of the heliospheric part ($0.1$~AU) for two HSS cases (one during a period of low and one during a period of high solar activity). In section 5, the simulation results in the vicinity of Earth are presented and compared, while in Section 6, we discuss the conclusions and possible future steps.


\section{The models}
\label{Section2}
\subsection{The default EUHFORIA set-up}

The modeling domain of EUHFORIA is divided in two distinct regions: the coronal part, which extends from the solar surface to $0.1$~AU, and the heliospheric part, which covers the spatial domain from $0.1$~AU till 2 AU \citep{pomoell18}. The coronal part provides the inner boundary conditions 
necessary for the initiation of the heliospheric part. In the default EUHFORIA set-up, the MHD wind parameters at $0.1$~AU are provided with 2 deg resolution by the semi-empirical Wang-Sheeley-Arge \citep[WSA;][]{arge03, Arge04} model, in combination with the potential field source surface (PFSS) model \citep{altschuler69, Wiegelmann2017} and the Schatten current sheet (SCS) model \citep{schatten69}. Using these initial boundary conditions in an MHD relaxation procedure, we obtain a steady heliospheric background wind. Then, CMEs are inserted into the background wind at $0.1$~AU, and their evolution and propagation throughout the heliosphere is modeled by solving the 3D time-dependent MHD equations, while taking into account interactions with the solar wind. The equations are solved in the Heliocentric Earth Equatorial (HEEQ) coordinate system, namely, in the system that has its Z axis parallel to the rotation axis of the Sun and its X axis towards the intersection of the solar equator and the solar central meridian as seen from the Earth.

\begin{figure*}
\centering
\begin{subfigure}[]{0.465\linewidth}
\includegraphics[width=\linewidth]{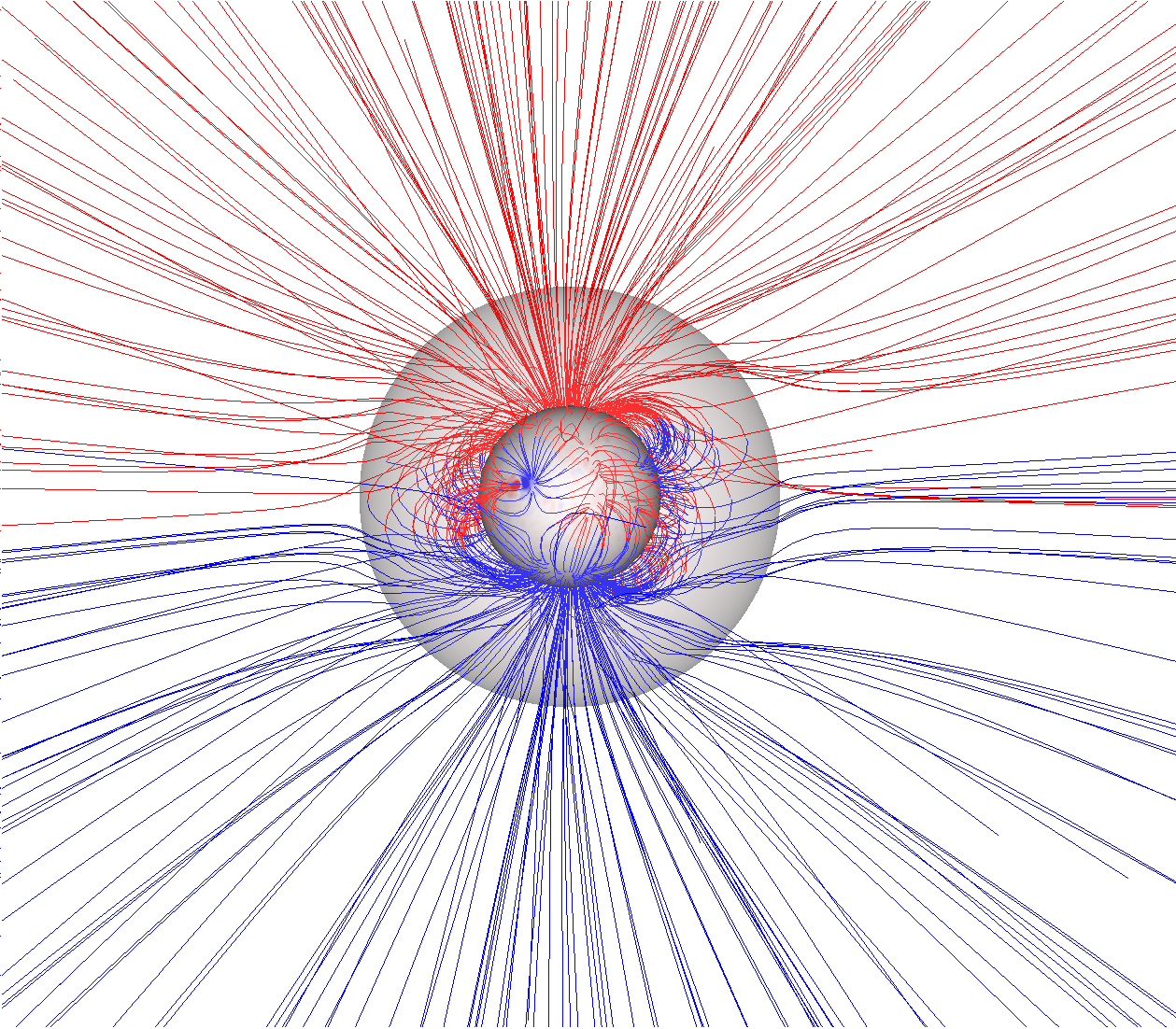}
\end{subfigure}
\hspace{0.01em}%
\begin{subfigure}[]{0.44\linewidth}
\includegraphics[width=\linewidth]{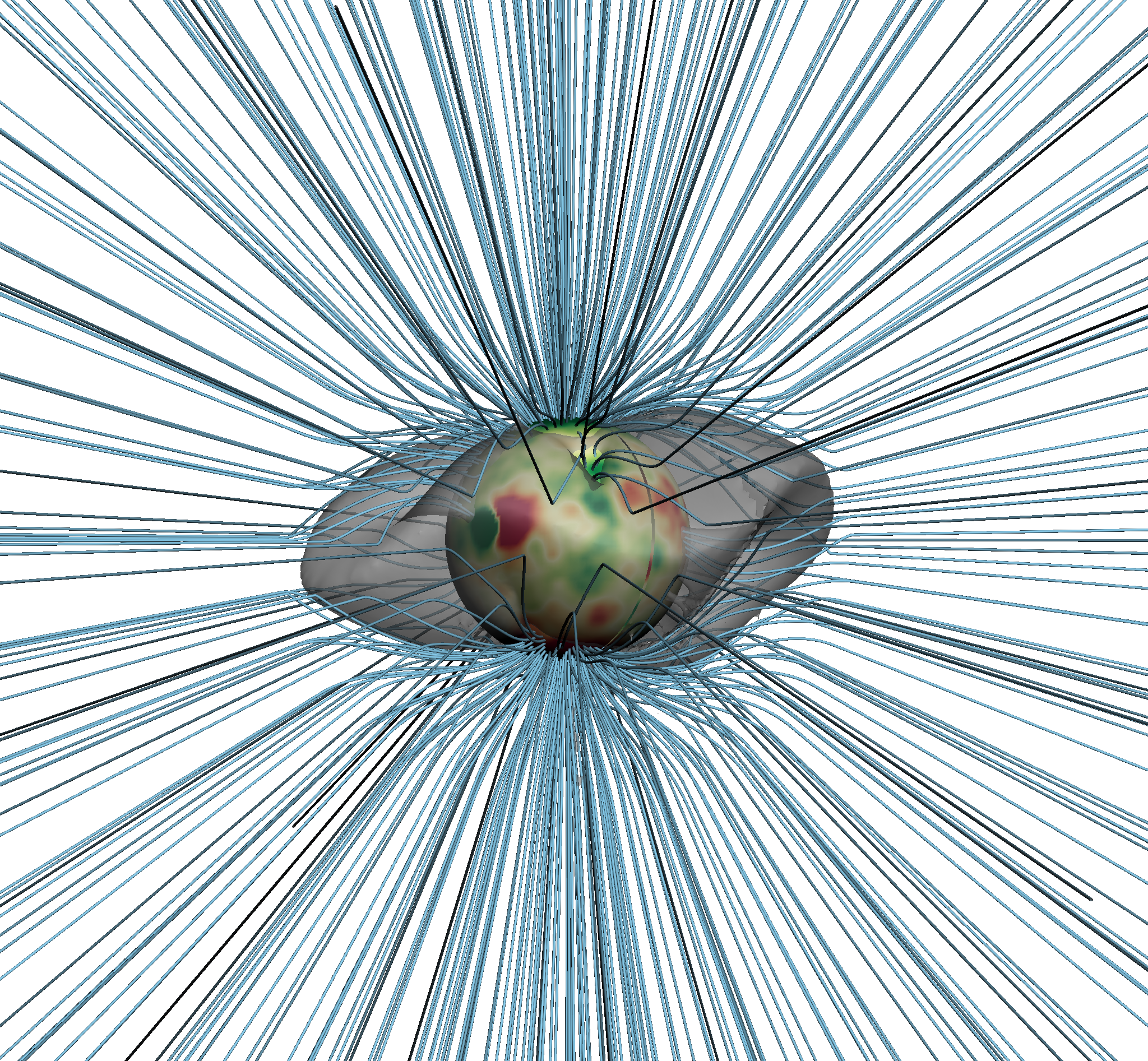}
\end{subfigure}%
\caption{Left: The coronal magnetic field as reconstructed by the PFSS and the SCS in EUHFORIA with a GONG ADAPT magnetogram on 2018-05-05T00:00. Red and blue colors indicate the opposite magnetic field polarity between the two solar hemispheres. The inner grey sphere depicts the solar photosphere while the outer transparent sphere marks the radius of $2.3\;$R$_{\odot}$, beyond which the SCS starts taking place. Right: The coronal magnetic field as reconstructed by MULTI-VP using the same magnetogram. The green and red colors on the solar surface represent the polarity of the magnetic field while the grey area represents the boundary between closed and open field lines.}
\label{Fig:EUHFORIA&MVP_models}
\end{figure*}

The first input to EUHFORIA's coronal part is the magnetogram that provides the necessary line-of-sight magnetic flux density information. Synoptic magnetograms provided by the Global Oscillation Network Group (GONG), the Helioseismic and Magnetic Imager (HMI) on board SDO spacecraft, and the Wilcox Space Observatory (WSO) or synchronic magnetograms such as the GONG Air Force Data Assimilative Photospheric Flux Transport Model (ADAPT) and HMI ADAPT, can be used as input to the model. After the magnetogram has been inserted and read, we employ the PFSS model until the height of 2.6~R$_{\odot}$, to reconstruct a current-free magnetic field (see Fig.~\ref{Fig:EUHFORIA&MVP_models}a). 

Starting from the height of $2.3\;$R$_{\odot}$ onwards, we employ the SCS model (see Fig.~\ref{Fig:EUHFORIA&MVP_models}a). This model starts before the end of the PFSS domain in order to reduce possible kinks in the magnetic field lines due to incompatible boundary conditions between the two models \citep[see e.g.][]{mcgregor11, Asvestari19}. The purpose of the SCS is to create an approximately uniform coronal magnetic field away from the Sun, maintaining a thin structure for the heliospheric current sheet (HCS). A more uniform magnetic field is necessary in order to obtain a better agreement between the model and the observations. In particular, the Ulysses mission data suggest that the radial magnetic field component is invariant of latitude \citep{Balogh95,Pinto17}.

After the 3D magnetic field reconstruction up to $0.1$~AU is completed, the large-scale topology is determined. This allows identification of "open" and "closed" magnetic field lines (i.e., magnetic field lines originating from coronal holes or magnetic field lines that shape closed loops) and their connectivity to the photosphere. In order to determine which of them are found open/closed to the solar wind, a tracer follows every magnetic field line from the photosphere upwards. If the line continues high in the corona, it is marked as "open", while if it returns back to the photosphere, it is defined as "closed" \citep[see][]{pomoell18}. The connectivity of the magnetic field lines is determined by tracing the open magnetic field lines from the outer boundary of the coronal domain ($0.1$~AU) inwards, towards the photosphere. From this tracing, the flux tube expansion factor of each magnetic field line can be calculated based on the relation \citep[see][]{Riley15}: 

\begin{equation}
f= \left(\frac{R_{\odot}}{R_{b}}\right)^{2} \frac{B_{r}(R_{\odot}, \theta,\phi)}{B_{b}(R_{b}, \theta_{b}, \phi_{b})},
\label{fte_factor}
\end{equation}

\noindent where $R_{b}=0.1\;$AU while $B_{r}$ and $B_{b}$ are the radial magnetic field at the photosphere and at $0.1$~AU, respectively. The flux tube expansion factor provides a quantification of the rate at which a flux tube cross-section expands, from the photosphere up to $0.1$~AU, as compared to purely radial expansion \citep{Wang&Sheeley90}. 

The factors that mainly define the WSA velocity formula (in km/s) used in the default set-up of EUHFORIA are the minimum angular distance of the foot-point of every magnetic field line to the closest coronal hole boundary (\textit{d}) and the flux tube expansion factor (f): 

\begin{equation}
 v_{r}(f,d)=240 +\frac{675}{(1+f)^{0.222}}\left[1-0.8\exp\left(-\left(\frac{d}{0.02}\right)^{1.25}\right)\right]^3.
\label{WSA_vr}
\end{equation}

\noindent Equation~\ref{WSA_vr} is a semi-empirical equation that provides the radial velocity of the solar wind, $v_{r}$, at the inner boundary of the heliospheric domain \citep{vanderHolst10, mcgregor11, pomoell18}. Based on this relation, the density ($n$), temperature ($T$) and radial magnetic field ($B_{r}$) are calculated at the boundary, as follows:

\begin{equation}
n=n_\textup{fsw}(v_\textup{fsw}/v_{r})^{2},
\label{WSA_n}
\end{equation}

\begin{equation}
    T=T_\textup{fsw}(\rho_\textup{fsw}/\rho),
\label{WSA_T}
\end{equation}
and
\begin{equation}
    B_{r}=\textup{sgn}(B_\textup{corona})B_\textup{fsw}(v_{r}/v_\textup{fsw}).
\label{WSA_Br}
\end{equation}

\noindent where v$_\textup{fsw}$ = 675 km/s is the velocity of the fast solar wind that carries a magnetic field of B$_\textup{fsw}$ = 300 nT at 0.1 AU.
The plasma number density of the fast solar wind at the same radius is n$_\textup{fsw}$ = 300 cm$^{-3}$ while \textup{sgn}(B$_\textup{corona}$) is the sign of the magnetic field as given by the coronal model. Also, the plasma thermal pressure is constant at the boundary and equal to P = 3.3 nPa corresponding to a temperature of T$_\textup{fsw}$ = 0.8 MK in the fast solar wind (see \citet[][]{pomoell18} for more details). The parameter $\rho$ denotes the mass density with $\rho_\textup{fsw}$ = 0.5n$_\textup{fsw}$m$_{p}$, where m$_{p}$ is the proton mass. Note that the estimation of the radial magnetic field at $0.1$~AU is not obtained directly from the reconstructed magnetic field, but it is re-calculated based on the empirical WSA velocity of Eq.~\ref{WSA_vr}. By employing this technique, we avoid the "open flux problem" \citep{linker17}, i.e.\ the problem of the magnetic field strength underestimation in interplanetary space, when inferred from coronal models. We underline this point, as it will be important for the continuation of this analysis. Moreover, for the remainder of this study, the default EUHFORIA coronal model will be referred to as the "WSA$^*$" model keeping in mind that a version of WSA is used in combination with the PFSS and SCS models. 

\subsection{The MULTI-VP model}

The MULTI-VP model \citep{Pinto17} is a coronal model that solves a set of equations (eq. \ref{MVP_1}, \ref{MVP_2}, \ref{MVP_3}) to provide the solar wind conditions at $0.1$~AU (wind speed, density, temperature and magnetic field) based on the following equations:

\begin{equation}
    \partial_{t}\rho + \nabla \cdot(\rho \text{u}) = 0,
\label{MVP_1}
\end{equation}

\begin{equation}
    \partial_{t}\text{u}+(\text{u}\cdot\nabla_{s})\text{u}=-\frac{\nabla_{s}P}{\rho}-\frac{GM}{r^{2}}\cos\alpha +v\nabla_{s}^{2}u,
\label{MVP_2}
\end{equation}

\begin{equation}
    \partial_{t}T+\text{u}\cdot\nabla_{s}T+(\gamma-1)T\nabla\cdot \text{u}=-\frac{\gamma-1}{\rho} [\nabla \cdot F_\textup{h}+\nabla\cdot F_\textup{c}+\rho^{2}\Lambda(T)].
\label{MVP_3}
\end{equation}

\noindent These equations describe a guided solar wind flow, constrained by the geometry of the magnetic flux-tubes that drives it in the low plasma beta limit. Many contiguous 1D solar wind solutions that describe the heating and acceleration of a wind stream along individual magnetic flux-tubes are computed ($\rho$ is the mass density, u is the field-aligned wind speed and $T$ the plasma temperature). To keep the presentation simpler, the magnetic field does not appear explicitly in these equations but it is implicit in the definition of the $\alpha$ angle, and on the co-linear gradient and divergence operators for a 1D non-spherically expanding flow. The $\alpha$ angle denotes the angle between the magnetic field and the radial inclination in respect to the vertical direction. The co-linear gradient depends on the magnetic field and its gradient ($B$ and grad($B$)). Energy and momentum transport along the field are fully taken into account. The individual solar wind profiles are computed on a grid of points aligned with the magnetic field, and therefore $\nabla_s$ represents derivatives along the magnetic field direction. The parameter $r$ represents the radial coordinate. The ratio of specific heats is $\gamma = 5/3$. The terms $F_\textup{h}$ and $F_\textup{c}$ denote the mechanical heating flux and the Spitzer-H\"arm conductive heat flux, which are both field-aligned. The term $\Lambda(T)$ denotes the radiative loss rate (see \citet{Pinto17} for more details). $F_\textup{h}$, $F_\textup{c}$ and $\Lambda(T)$ are fixed according to the calibrations done in the model and do not vary throughout the simulations.
The 1D solutions, altogether, sample the whole solar atmosphere (or any sub-domain of interest). As in WSA$^*$, the coronal field topology is the main external constraint: a magnetogram is used as the first source of information and PFSS extrapolations undertake the task of reconstructing a current-free magnetic field, up to the distance of $2.5$ R$_{\odot}$ (see Fig.~\ref{Fig:EUHFORIA&MVP_models}b). 
The MULTI-VP model does not use, though, the SCS model to achieve uniformity of the magnetic field in the high corona. Instead, it applies a flux-tube expansion profile that smoothly and asymptotically transforms the non-uniform field at the source surface, into a uniform field at $\approx12\;$R$_{\odot}$ by conserving the total open magnetic flux. The radial magnetic field at the outer boundary is provided directly by the corrected PFSS extrapolations, in contrast to the radial magnetic field obtained by WSA$^*$, which is calculated based on Eq.~\ref{WSA_Br}, as explained earlier. 


\section{Interfacing MULTI-VP and EUHFORIA-heliosphere}

MULTI-VP provides directly the full set of physical quantities required by EUHFORIA-heliosphere as boundary conditions. We hence set up MULTI-VP to compute the full spherical domain (all latitudes and longitudes) at standard angular resolution of $2$ degrees (for GONG magnetogams) or $5$ degrees (for WSO magnetograms), and transmit maps of $v_r$, $n$, $T$ and $B_r$ computed at $0.1$~AU to EUHFORIA-heliosphere.
The interfacing procedure undergoes an intermediate verification step to confirm the general validity of the inputs, and specifically to ensure that the wind is super-critical everywhere at the interface between the two models. EUHFORIA-heliosphere requires by construction that this condition is met.
Sub-critical speeds at the inlet boundary would otherwise lead to erroneous modeling results.
However, coronal models can occasionally produce wind flows that are sub-critical at $0.1$~AU. These can correspond, for example, to very slow wind streams that have not reached their asymptotic state (that is, that have not yet accelerated to super-critical speeds) at that altitude.
Spurious features and irregularities in the extrapolated magnetic fields can also produce errors.
Hence, we consistently search for sub-critical wind speeds at $0.1$~AU, and apply corrections where needed. 
Our approach consists of adjusting each individual sub-critical wind stream by increasing its speed while decreasing its density, conserving its mass flux. This approximately corresponds to "pushing" the wind flow closer to its asymptotic state by mimicking the effect of a spatially extended acceleration region. The target wind speed is estimated from the speed values of the adjacent streams. The solar wind speed is further clipped at reasonable lower and upper limits to remove outliers, and density is recalculated to conserve the mass flux.

\begin{figure}[h!]
     \centering
     \begin{subfigure}[b]{0.5\textwidth}
         \centering
         \includegraphics[width=\textwidth]{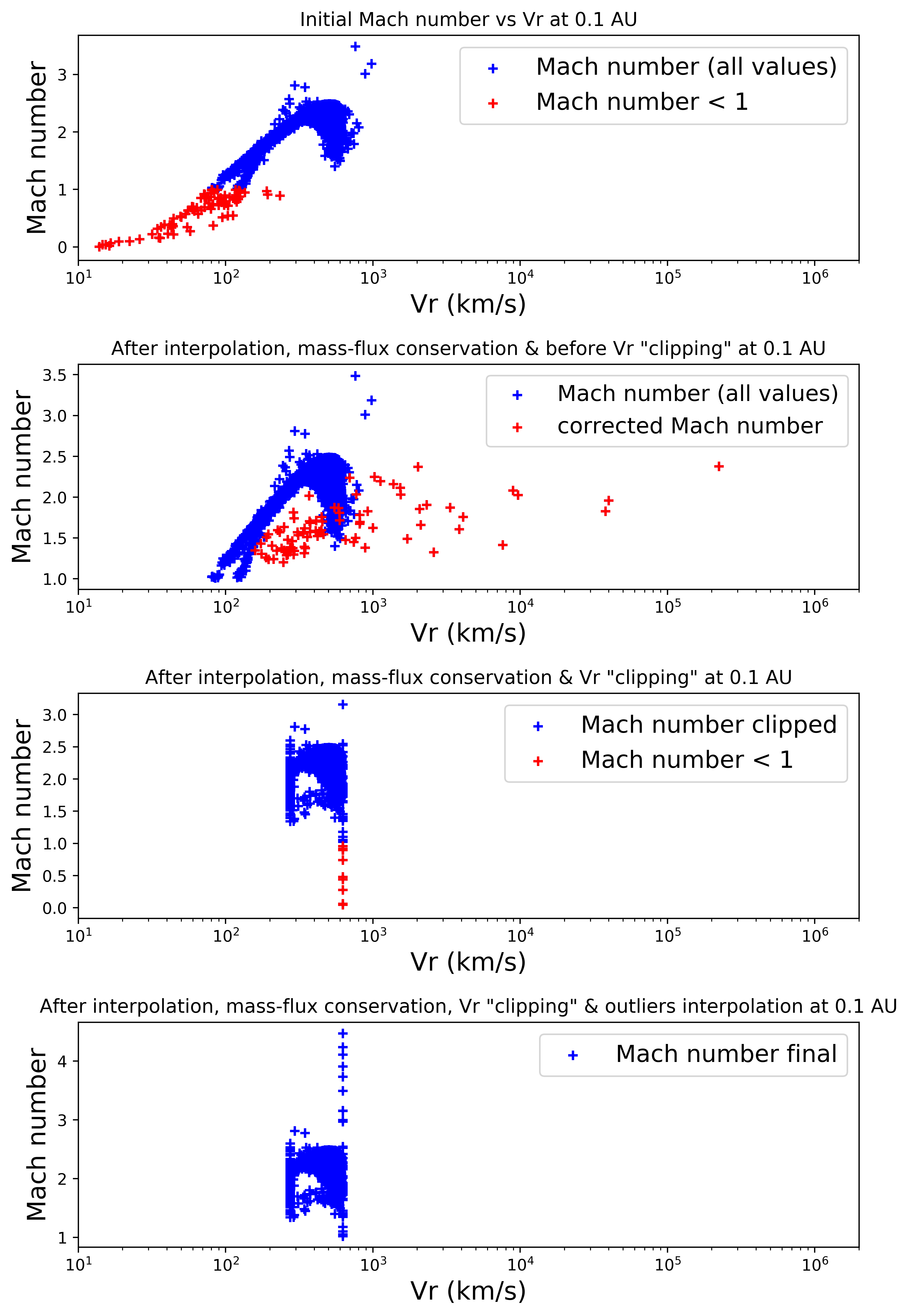}
     \end{subfigure}
     \caption{Fast magnetosonic Mach number as a function of radial velocity at $0.1$~AU. The different panels indicate the steps followed to correct the sub-critical values from the MULTI-VP model before inserting them into the heliospheric part of EUHFORIA.}
     \label{Fig:SubalfvenicCorrections_mvp}
\end{figure} 

The applied procedure is illustrated in Fig.~\ref{Fig:SubalfvenicCorrections_mvp} and discussed in detail below:

\begin{enumerate}
 \item 
We locate all pixels of the interface solar wind maps with a sub-critical fast magnetosonic Mach number ($M) < 1$ (first panel of Fig.~\ref{Fig:SubalfvenicCorrections_mvp}).

\item 
The Mach numbers of such pixels are bilinearly interpolated with their first and closest super-critical neighbors. In case one of the neighboring pixels is also sub-critical, it is ignored during the interpolation. In case there are many sub-critical pixels grouped together (horizontally or vertically), the first pixel along the direction of interpolation is corrected based on the aforementioned procedure while the rest of the pixels (2nd, 3rd pixel, etc) take into account the new, super-critical value of the previously sub-critical pixel for the continuation of interpolation.

\item 
Based on the new Mach numbers, we calculate the new radial velocity $v_r$ and density $n$ at the boundary by conserving the mass-flux (second panel of Fig.~\ref{Fig:SubalfvenicCorrections_mvp}).

\item 
We furthermore restrict the radial velocity at the boundary between 275 and $625\;$km/s in order to comply with the scale that is applied to WSA (see \citet{mcgregor11}; and default EUHFORIA set-up, \citet{pomoell18}). Then, we employ again the mass-flux conservation to calculate the new $n$ value. This last step can occasionally still lead to a few pixels with $M < 1$ (third panel of Fig.~ \ref{Fig:SubalfvenicCorrections_mvp}) due to the significantly low estimated densities (only a few cm$^{-3}$). These pixels are assumed as "deviations" or "outliers" and we interpolate their low densities with the ones from their closest super-critical neighbors in order to achieve $M > 1$ (see last panel of Fig.~\ref{Fig:SubalfvenicCorrections_mvp}).

\end{enumerate}


\section{Comparison between MULTI-VP and WSA$^*$ results at $0.1$~AU}

The analysis presented below is focused on two different HSS events. One during a period of low solar activity (HSS reached Earth on 2018-01-21) and another during a period of high solar activity (HSS reached Earth on 2011-06-22). These events were selected because the default set-up of EUHFORIA did not produce accurate results. Replacing WSA$^{*}$ with MULTI-VP, could help us understand how the difference in coronal models affects the final simulation results at Earth. We further investigate how our results differ due to the use of different magnetograms (GONG and WSO) for the two distinct model set-ups.

\subsection{HSS case during the period of low solar activity}
\subsubsection{Results with GONG magnetograms}

\begin{figure}[ht]
\centering
\begin{subfigure}[b]{0.48\textwidth}
   \includegraphics[width=1\linewidth]{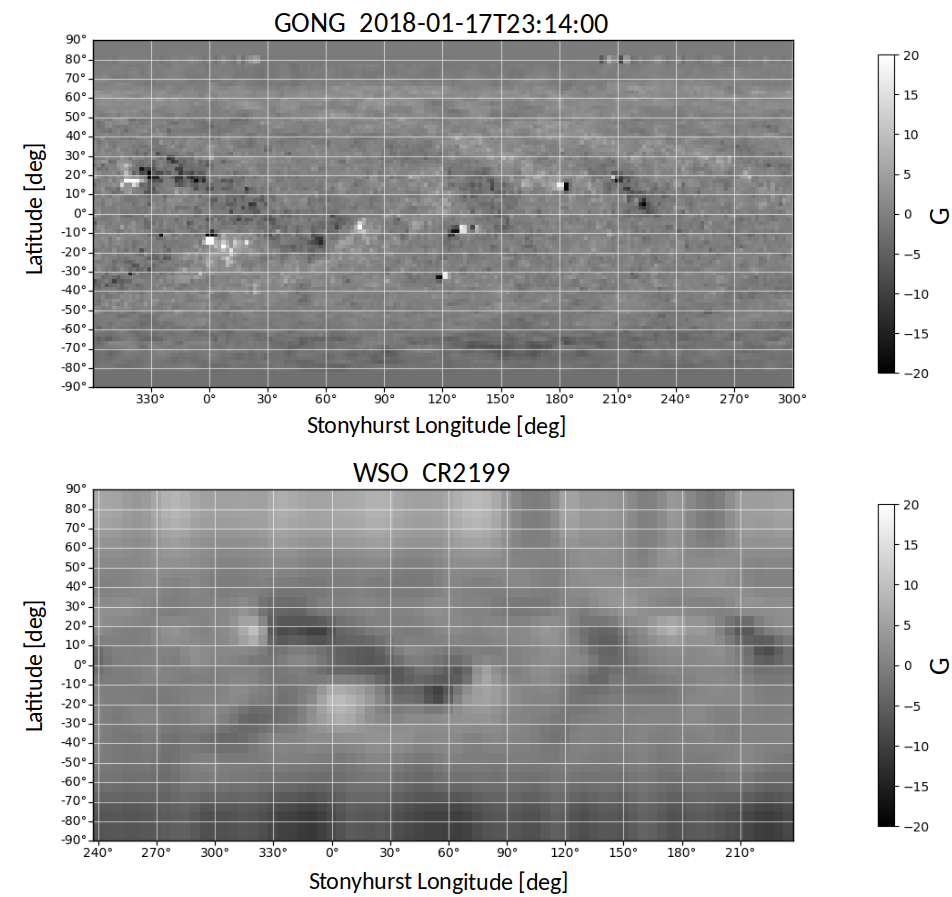}
\end{subfigure}
\caption{GONG (upper panel) and WSO (lower panel) magnetograms that were used for the HSS case during the period of low solar activity.}
\label{Fig:magnetograms_sola_min}
\end{figure}

\begin{figure*}
\centering
\begin{subfigure}[b]{0.85\textwidth}
   \includegraphics[width=1\linewidth]{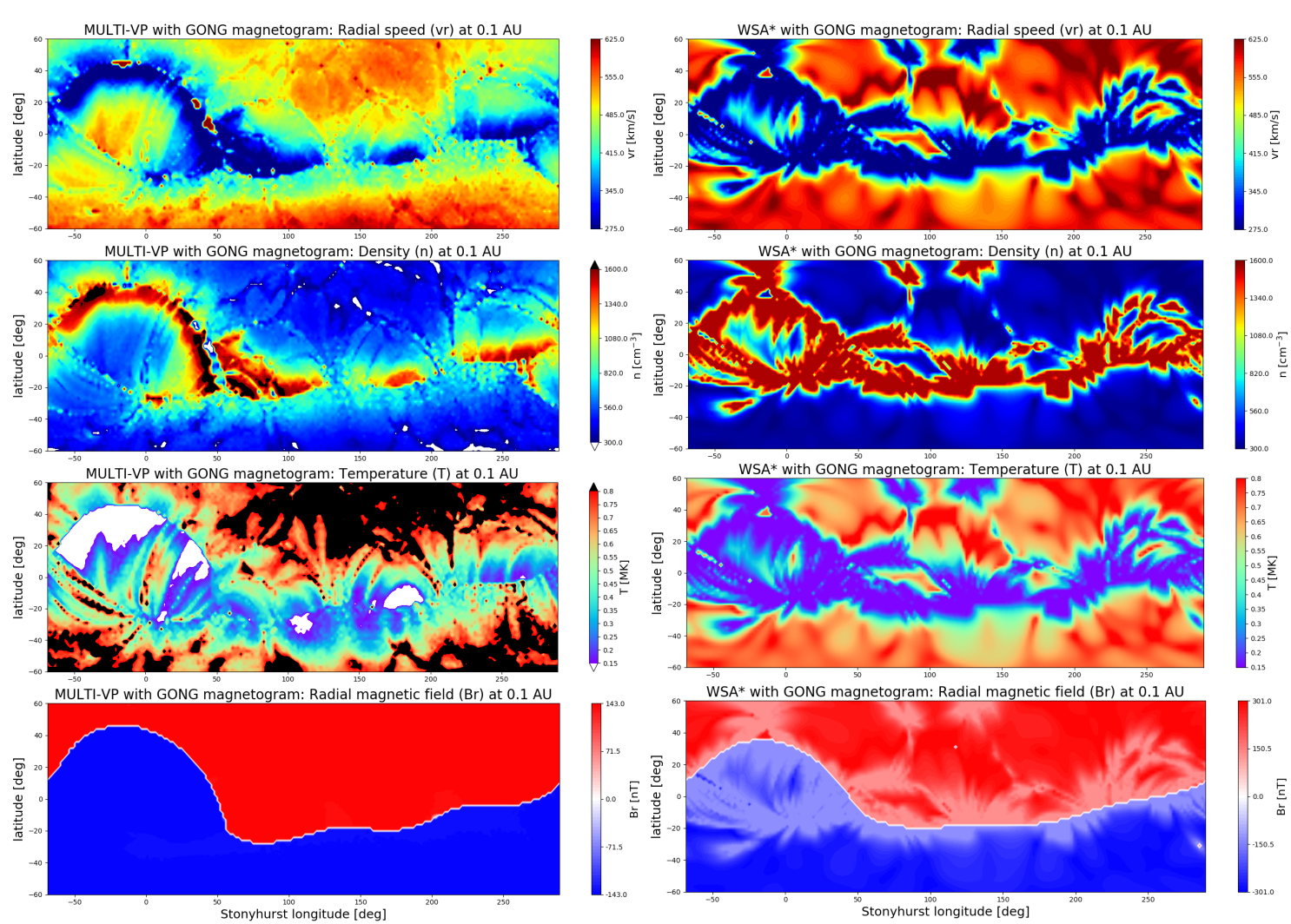}
   \caption{}
   \label{fig:Ng1} 
\end{subfigure}

\begin{subfigure}[b]{0.85\textwidth}
   \includegraphics[width=1\linewidth]{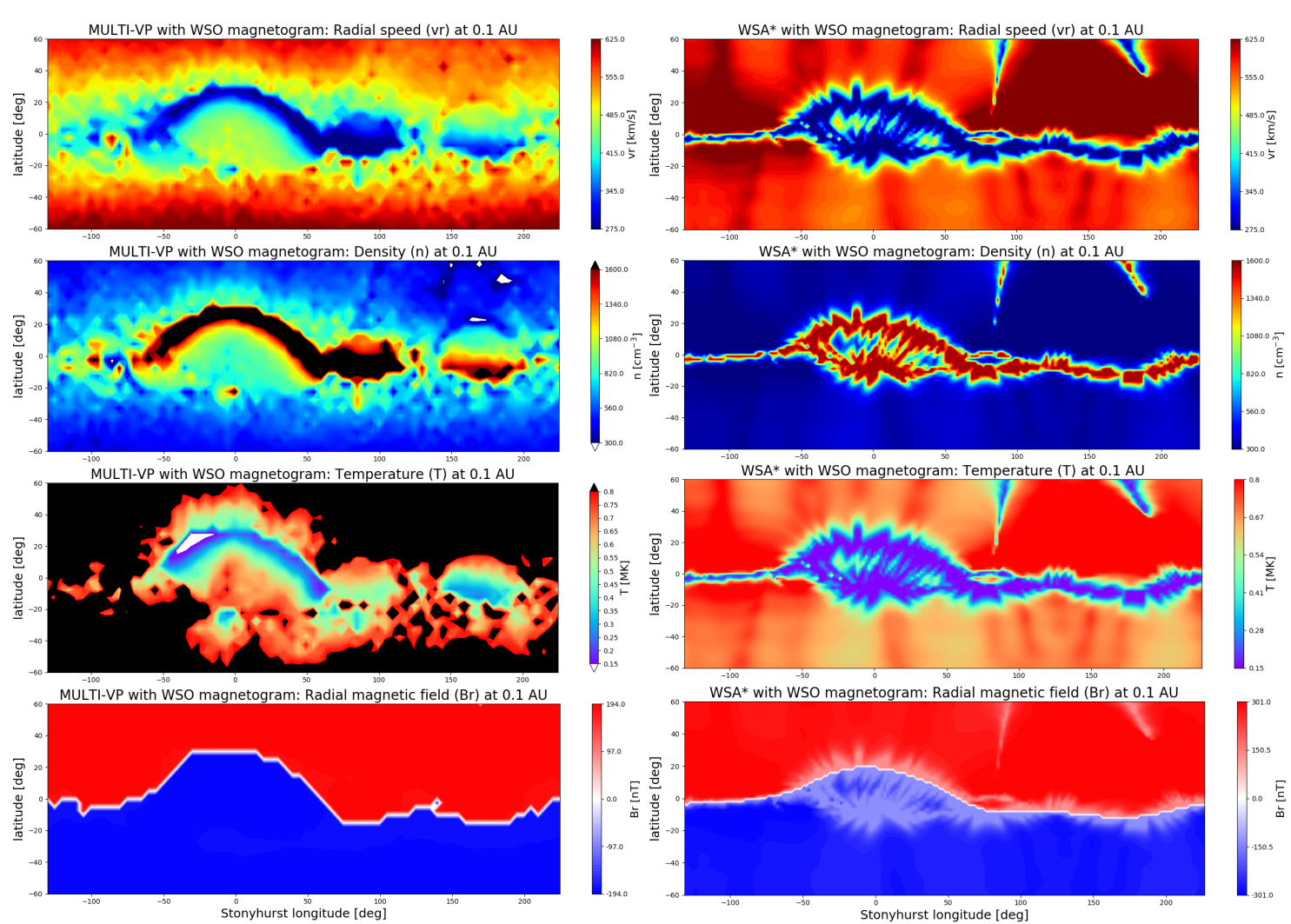}
   \caption{}
   \label{fig:Ng2}
\end{subfigure}

\caption{Latitude-longitude maps of radial velocity ($v_{r}$), density ($n$), temperature ($T$) and radial magnetic field ($B_{r}$) as obtained by MULTI-VP (left column) and WSA$^*$ (right column) at $0.1$~AU with a GONG (a) and a WSO (b) synoptic magnetogram for the HSS case during the period of low solar activity.}
\label{Fig:2Dmaps_solar_min}
\end{figure*}

The upper panel of Fig.~\ref{Fig:magnetograms_sola_min} shows the GONG synoptic magnetogram in Stonyhurst coordinates \citep{Stonyhurst} used as input to both MULTI-VP and WSA$^*$, for the modeling of the HSS that arrived at Earth on 2018-01-21. In Fig.~\ref{Fig:2Dmaps_solar_min}a the resulting boundary conditions at $0.1$~AU are plotted as latitude-longitude maps. The left column presents the radial velocity ($v_r$), particle density ($n$), temperature ($T$) and radial magnetic field ($B_r$) as given by MULTI-VP while the right column presents the same quantities, as produced by WSA$^*$.

\begin{figure*}
\centering

\begin{subfigure}[c]{0.47\linewidth}
\includegraphics[width=\linewidth]{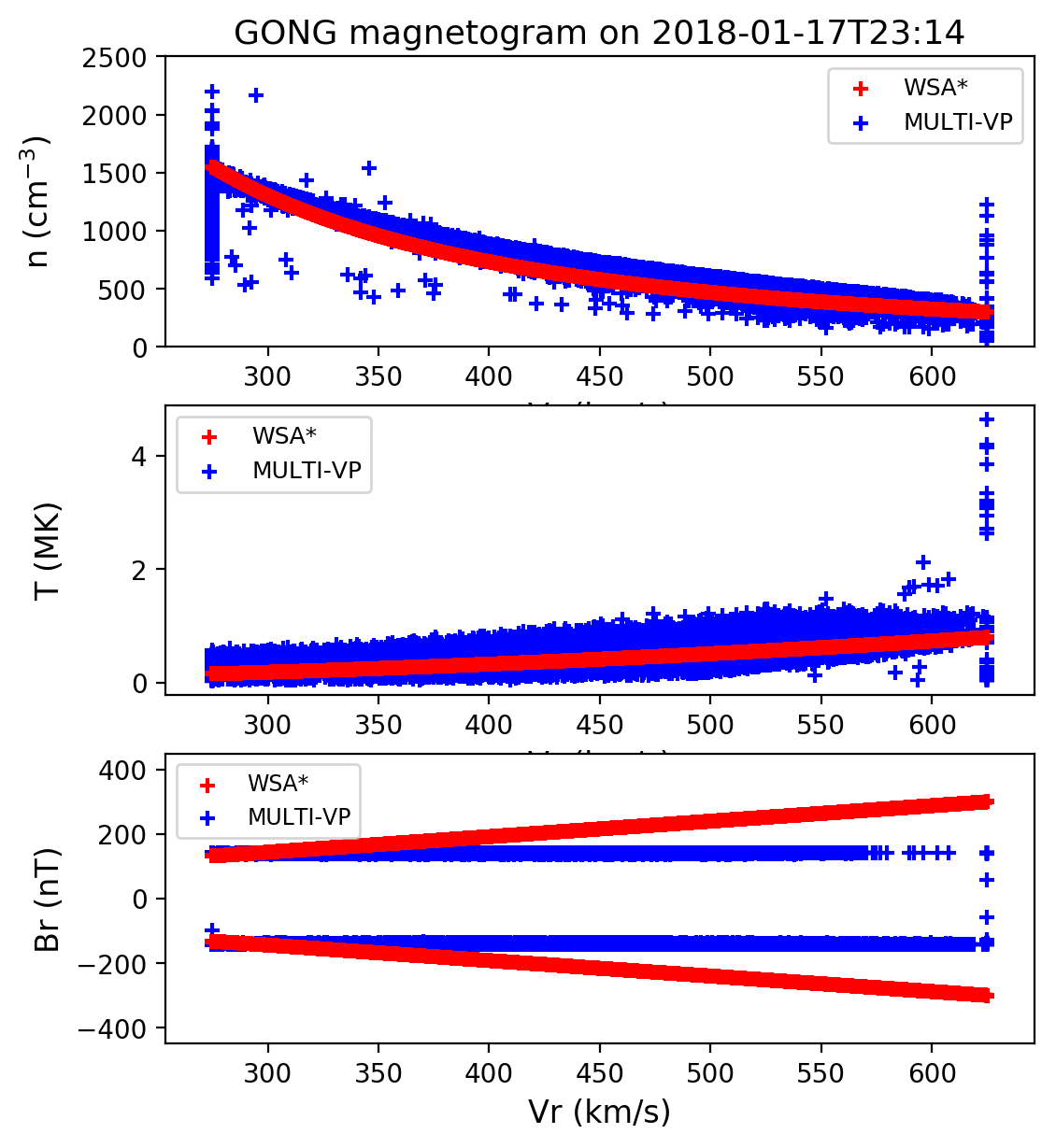}
\caption{}
\end{subfigure}
\hspace{0.1em}%
\begin{subfigure}[d]{0.47\linewidth}
\includegraphics[width=\linewidth]{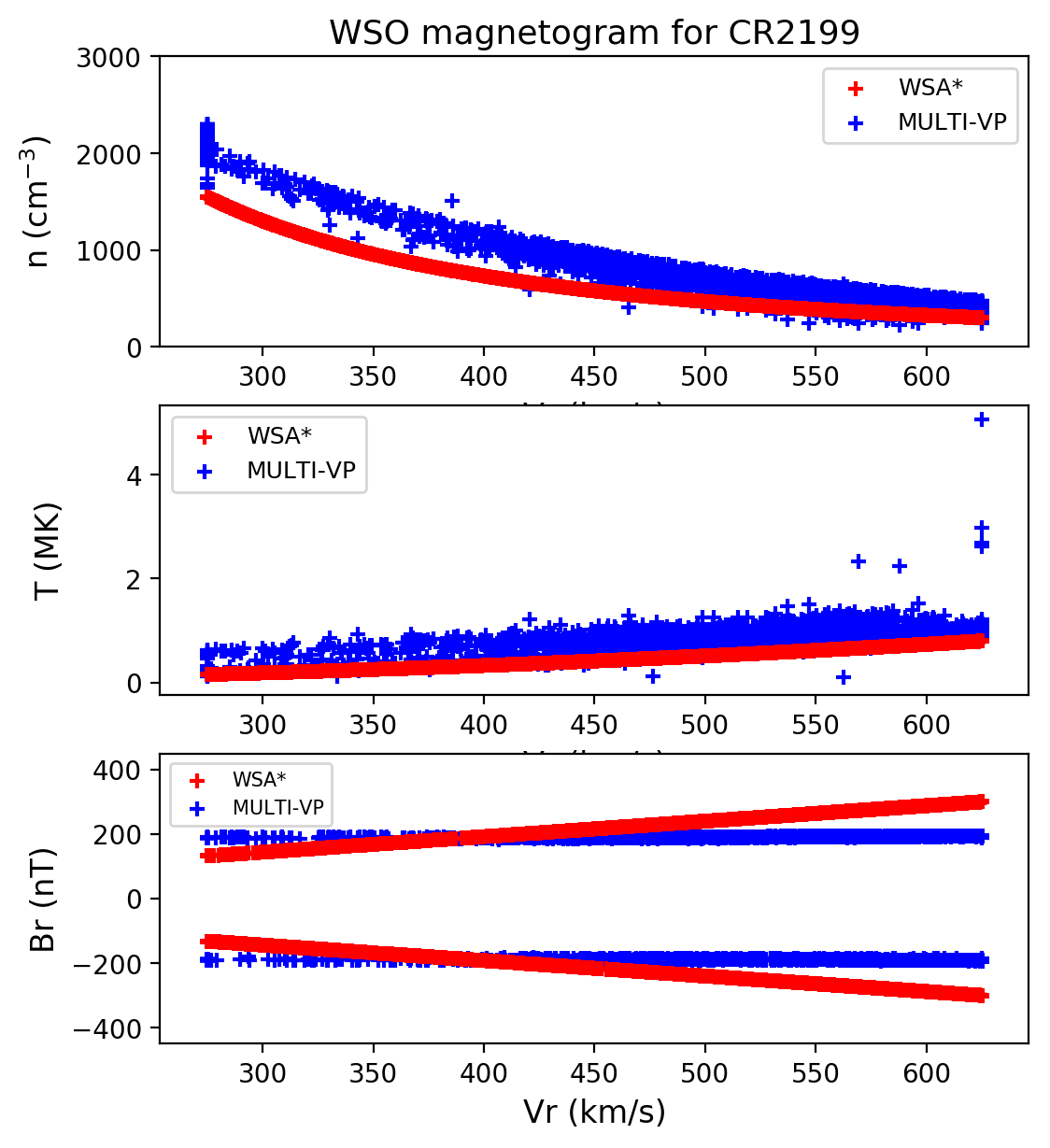}
\caption{}
\end{subfigure}%

\begin{subfigure}[c]{0.47\linewidth}
\includegraphics[width=\linewidth]{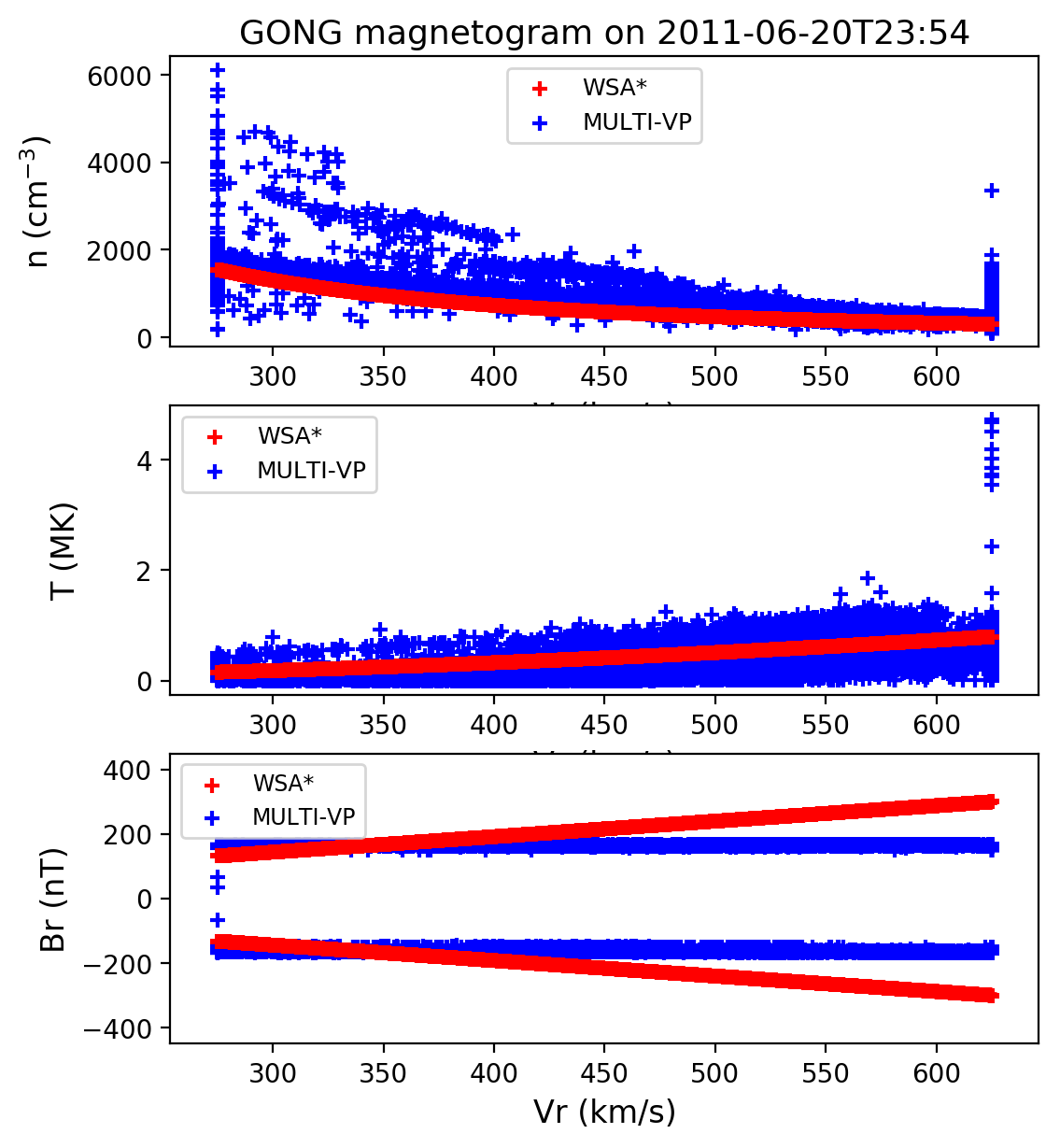}
\caption{}
\end{subfigure}
\hspace{0.1em}%
\begin{subfigure}[d]{0.47\linewidth}
\includegraphics[width=\linewidth]{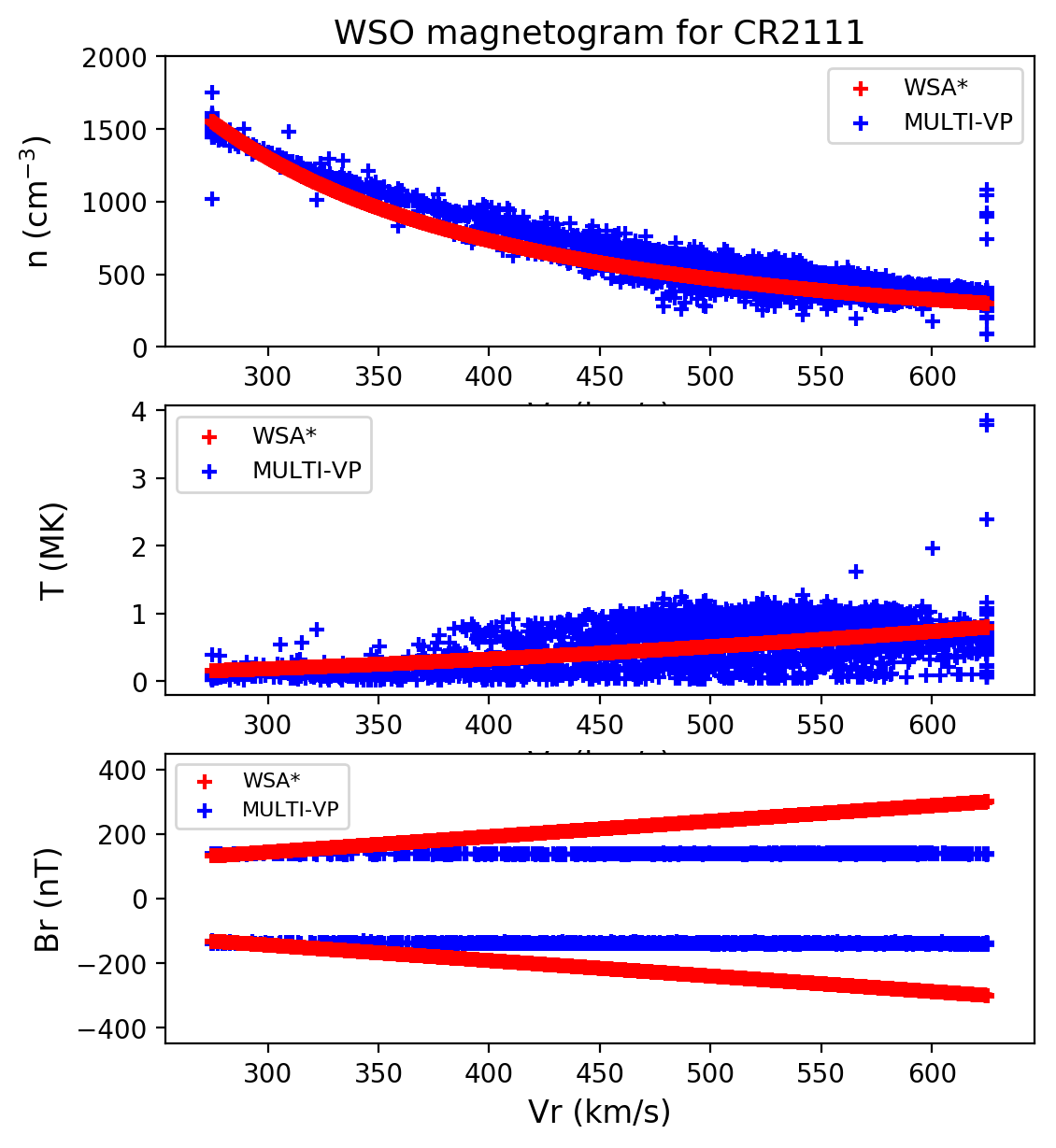}
\caption{}
\end{subfigure}%

\caption{The $n-v_{r}$, $T-v_{r}$, $B_{r}$-$v_{r}$ distributions for both models and magnetograms at $0.1$~AU.}
\label{Fig:Boundary_parameters}
\end{figure*}

A number of differences can be observed in the maps produced by the two models. The radial velocity maps show the coronal hole (CH) region from which the studied HSS originated, extending between $[-50,0]^{\circ}$ in longitude and $\approx[-20,20]^{\circ}$ in latitude. The solar wind emerging from that CH is faster for MULTI-VP than for WSA$^*$. Overall, the WSA$^*$ model provides higher velocities for latitudes above $20^{\circ}$ and below $-30^{\circ}$ compared to MULTI-VP. Another significant difference is the distribution of lower solar wind speeds around the HCS. In the WSA$^*$ case, we distinguish a wider zone of slow wind than in the MULTI-VP maps. The regions of low speeds expand not only along the HCS zone but also towards the north pole, further surrounding the CH area below the equator. Figure~\ref{Fig:2Dmaps_solar_min}a further shows that MULTI-VP produces (a) higher densities and lower temperatures around the HCS compared to WSA$^*$, and (b) lower densities and higher temperatures for latitudes above $\approx20^{\circ}$ and below $\approx-20^{\circ}$ (black/white areas). The lower left panel ($B_{r}$) of Fig.~\ref{Fig:2Dmaps_solar_min}a indicates that in the region extending between [-50, 50]$^{\circ}$ around the central meridian (CM), MULTI-VP provides a HCS that is inclined to higher latitudes compared to the one modeled by WSA$^*$. The difference arises because of the influence of the SCS model in the latter case, which tends to flatten the HCS above the source surface.
Another direct difference between the $B_{r}$ maps is the gradient produced by the WSA$^*$, but not by MULTI-VP. This gradient appears because WSA$^*$ "re-calculates" the radial magnetic field at the boundary, based on the empirical velocity (Eq.~\ref{WSA_Br}). Thus, the range of $B_{r}$ values is broad. On the other hand, the radial magnetic field at $0.1$~AU produced by MULTI-VP is obtained by the PFSS extrapolations and gets corrected by an additional expansion profile that is applied to make the field uniform. As a result, $B_{r}$ converges towards two values (one positive and one negative due to polarity change), as shown in Fig.~\ref{Fig:Boundary_parameters}a (third panel).

\begin{figure*}
\centering
\begin{subfigure}[b]{0.85\textwidth}
   \includegraphics[width=1\linewidth]{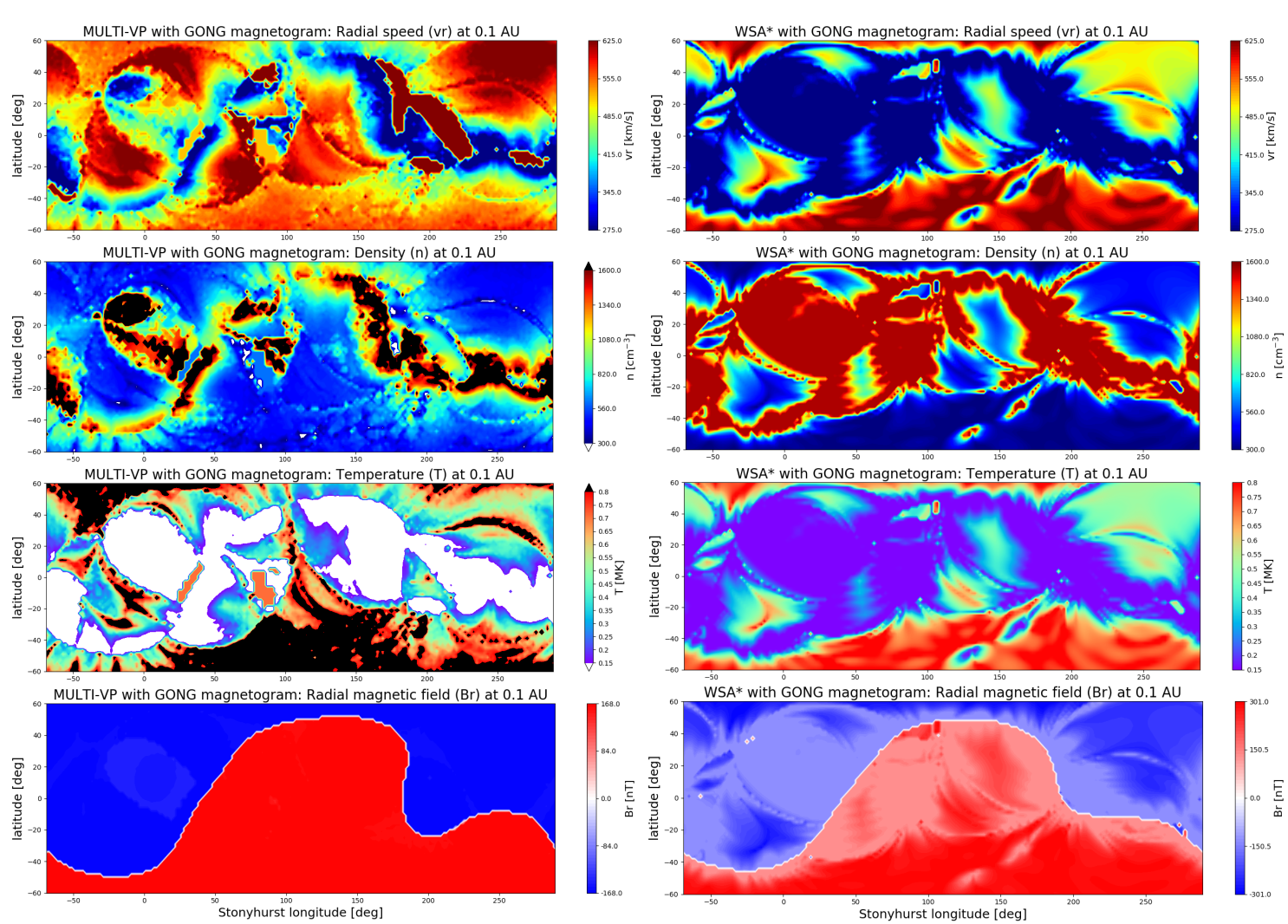}
   \caption{}
   \label{fig:Ng1} 
\end{subfigure}

\begin{subfigure}[b]{0.85\textwidth}
   \includegraphics[width=1\linewidth]{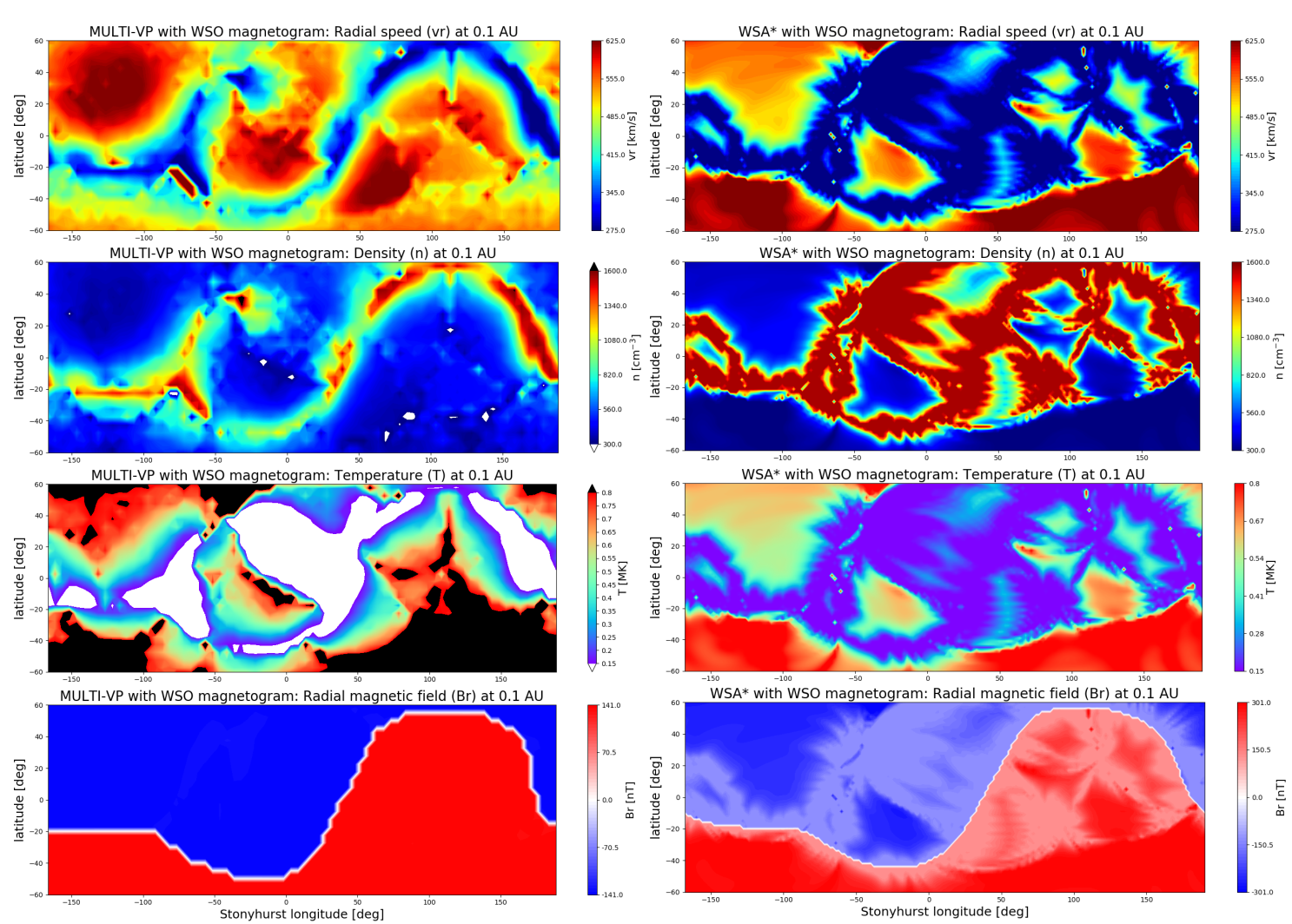}
   \caption{}
   \label{fig:Ng2}
\end{subfigure}
\caption{Latitude-longitude maps of radial velocity ($v_{r}$), density ($n$), temperature ($T$) and radial magnetic field ($B_{r}$) as obtained by MULTI-VP (left column) and WSA$^*$ (right column) at $0.1$~AU with a GONG (a) and a WSO (b) synoptic magnetogram, for the HSS case during the period of high solar activity.}
\label{Fig:2Dmaps_solar_max}
\end{figure*}

\subsubsection{Results with WSO magnetograms} 

Figure~\ref{Fig:2Dmaps_solar_min}b shows results for the HSS case during the period of low solar activity obtained by employing as input a WSO magnetogram (shown in the lower panel of Fig.~\ref{Fig:magnetograms_sola_min}). The WSO synoptic maps are not daily updated, contrary to GONG magnetograms, but produced for each Carrington rotation (CR). The date of our interest is included in the CR2199, thus, we used the corresponding magnetogram. 

In Fig.~\ref{Fig:2Dmaps_solar_min}b ($vr$ maps), the area between [-50, 0]$^{\circ}$ in longitude and $\approx [-20, 20]^{\circ}$ in latitude (which includes the CH from which the HSS of interest originated) is characterized by faster velocities in the MULTI-VP case, compared to the WSA$^*$ one.
The later map is characterized by higher velocities immediately above and below the HCS, compared to MULTI-VP that only provides such high velocities in regions above $\approx 30^{\circ}$ and below $\approx -30^{\circ}$ in latitude. Density and temperature from MULTI-VP show the same behavior as in the previous (GONG) HSS case during the low solar activity, i.e.,\ higher densities and lower temperatures around the HCS and lower densities and higher temperatures towards the poles, compared to WSA$^*$ (see white/black regions). The HCS is inclined again to higher latitudes for the MULTI-VP model, which does not produce any gradient compared to WSA$^*$ (lower panels of Fig.~\ref{Fig:2Dmaps_solar_min}b).

\subsubsection{Comparison of boundary results for different magnetograms} 

The comparison between Figs.~\ref{Fig:2Dmaps_solar_min}a and b shows a number of differences arising because of the different magnetograms. First, the GONG magnetograms have significantly higher resolution than the WSO ($1^{\circ}$ and $5^{\circ}$, respectively). GONG data are mapped onto a 2$^{\circ}$ x 2$^{\circ}$ grid at 0.1 AU while WSO on a 5$^{\circ}$ x 5$^{\circ}$ grid, for MULTI-VP, and a 2$^{\circ}$ x 2$^{\circ}$ grid for WSA$^*$. A second difference is directly seen in the extent of the HCS. Both the MULTI-VP and WSA$^*$ models show that the extent of the HCS is inclined to higher latitudes when GONG magnetograms are employed, compared to WSO. Furthermore, the slow solar wind around the HCS is more restricted to latitudes around the equator when using WSO magnetograms, compared to GONG.

Figure~\ref{Fig:Boundary_parameters} summarizes the $n-v_{r}$, $T-v_{r}$, $B_{r}$-$v_{r}$ distributions for both models at $0.1$~AU. This plot gives an overall idea of the differences between the two models, regarding the range of values and their amplitudes. The MULTI-VP boundary data are characterized by a range of density and temperature values for each velocity point. In the WSA$^*$ case, though, every velocity value corresponds to a single point in density and temperature. The wide range of densities at $\approx$ 275 km/s and 625 km/s in the MULTI-VP case, is due to the clipping of the velocities and the interpolation in densities, as explained in Section~3 (see also Fig.~\ref{Fig:SubalfvenicCorrections_mvp}). We remind the reader that the clipping was meant to keep the wind speed range consistent with that of WSA$^*$, and any resulting sub-critical values were further corrected.

\subsection{HSS case during the period of high solar activity}
\subsubsection{Results with GONG magnetogram}

\begin{figure}[h]
\centering
\begin{subfigure}[b]{0.48\textwidth}
   \includegraphics[width=1\linewidth]{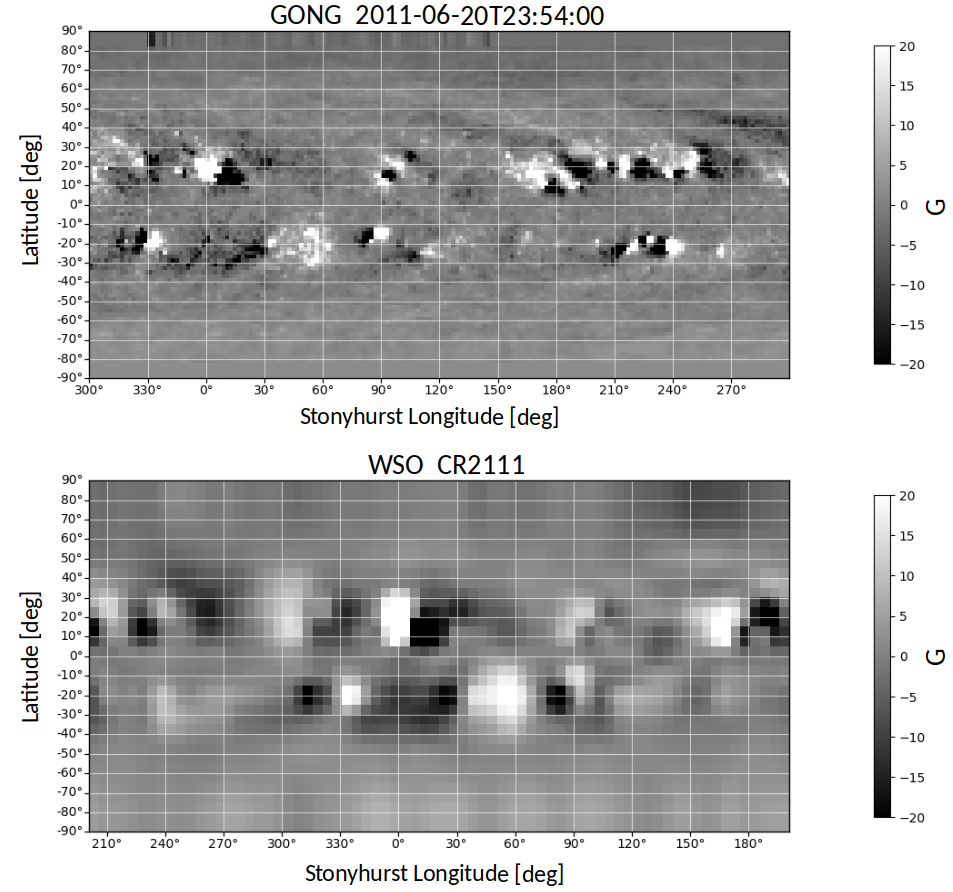}
\end{subfigure}
\caption{GONG (upper panel) and WSO (lower panel) magnetograms that were used for the HSS case during the period of high solar activity.}
\label{Fig:magnetograms_solar_max}
\end{figure}

The upper panel of Fig.~\ref{Fig:magnetograms_solar_max} shows the synoptic GONG magnetogram in Stonyhurst coordinates used as input to the coronal models for the simulation of the HSS case during the period of high solar activity. In Fig.~\ref{Fig:2Dmaps_solar_max}a, the obtained boundary conditions at $0.1$~AU are plotted as 2D maps, in the same coordinate system. The presented quantities are the same as in Fig.~\ref{Fig:2Dmaps_solar_min}a and b. 

The most distinctive characteristic of the MULTI-VP 2D velocity map (upper panel of Fig. \ref{Fig:2Dmaps_solar_max}a) is that it presents an elongated, fast velocity and low density, area between [170, 240]$^{\circ}$ in longitude. The WSA$^*$ model gives opposite results for the same region, i.e., dense area from which the slow solar wind emerges. We note that this was also the case for MULTI-VP before we applied the corrections for the sub-critical values. As mentioned in Section 3, a number of sub-critical values appeared in the MULTI-VP maps representing flows that did not manage to reach their asymptotic super-critical speeds at the altitude of 0.1 AU. After the corrections we imposed, the discussed region was transformed from a slow to a fast wind domain. This is because initially the specific area was composed by extremely low Mach numbers ($M$ < 1) which, after their interpolation and the mass-flux conservation, led to very high velocities (see Fig. \ref{Fig:SubalfvenicCorrections_mvp}, panel 2, velocities beyond 1000 km/s). These velocities were later clipped to the reasonable upper limit of 625 km/s (see Fig. \ref{Fig:SubalfvenicCorrections_mvp}, panel 3) and this is why the region of initially slow solar wind was transformed to an area of fast solar wind. For the HSS case we study here, the discussed area does not influence our results at Earth since it is situated at the back side of the Sun, as seen from Earth. 

Moreover, the temperatures obtained by MULTI-VP are higher than the ones modeled by WSA$^*$ for the areas around the poles, and lower for the regions around the HCS. The HCS in the MULTI-VP case extends again to higher latitudes compared to WSA$^*$, and as stated in the previous cases, it does not show any gradient (see discussion in section 4.1).

\subsubsection{Results with WSO magnetograms}

In Fig.~\ref{Fig:2Dmaps_solar_max}b, we present the 2D maps of plasma and magnetic parameters at the boundary for the HSS case during the period of high solar activity, based on the WSO magnetogram of CR2111 (lower panel of Fig.~\ref{Fig:magnetograms_solar_max}). The WSA$^*$ velocity, density and temperature maps show an extended slow, dense and cold solar wind region between $\approx[0,100]^{\circ}$ in longitude, which is not present in the MULTI-VP maps. We also observe that the CH located between $[-50, 0]^{\circ}$ in longitude seems more extended in the MULTI-VP than in the WSA$^*$ case. Faster solar wind is generated by MULTI-VP compared to WSA$^*$, originating from the extension of the northern polar CH found $\approx[-170,-70]^{\circ}$ in longitude. The same happens for the southern CH located below the equator ($\approx50^{\circ}$ in longitude) in the MULTI-VP map. More specifically, the latter, relatively "hot" fast wind area, is almost not present in WSA$^*$. Nevertheless, WSA$^*$ captures a second southern CH, located $\approx[100, 150]^{\circ}$ in longitude and at $\approx-20^{\circ}$ in latitude, which is not easily distinguished in the MULTI-VP case.
Besides the area of large discrepancy between $\approx[0,100]^{\circ}$ in longitude, the densities obtained by the two models qualitatively agree around the HCS, and also in the polar regions. On the other hand, the temperatures obtained by MULTI-VP are higher than the ones modeled by WSA$^*$ towards the poles, and lower around the HCS. The $B_{r}$ plots show a similar characteristic to the previous cases, where significant $B_{r}$ gradient is detected in WSA$^{*}$, but not in MULTI-VP. In this example, though, the HCS is inclined to approximately the same latitude, for both MULTI-VP and  WSA$^*$.

\subsubsection{Comparison of boundary results in the frame of different magnetograms} 

Figures ~\ref{Fig:2Dmaps_solar_max}a and \ref{Fig:2Dmaps_solar_max}b show that the WSA$^*$ results for both GONG and WSO magnetograms, contain an unexpected slow and dense solar wind region, visible in the range $[0,50]^{\circ}$ and $[0,100]^{\circ}$ in longitude, respectively. In the case of MULTI-VP with a GONG magnetogram, the elongated region of fast solar wind velocity between [170, 240]$^{\circ}$ in longitude disappears when using a WSO magnetogram indicating that it is not the coronal model or the magnetogram alone that produces distinctly different results. Thus, it is important to use an appropriate combination of input magnetogram and coronal model to achieve the optimal output. 

Furthermore, the $n-v_{r}$, $T-v_{r}$, $B_{r}$-$v_{r}$ distributions for both models are presented in Fig.~\ref{Fig:Boundary_parameters}c and d. Overall, the MULTI-VP boundary data are characterized by a range of density and temperature values for each velocity value, as also stated in subsection 4.1.3. 


\begin{figure*}[!ht]
     \centering
     \begin{subfigure}[b]{0.95\textwidth}
         \centering
         \includegraphics[width=\textwidth]{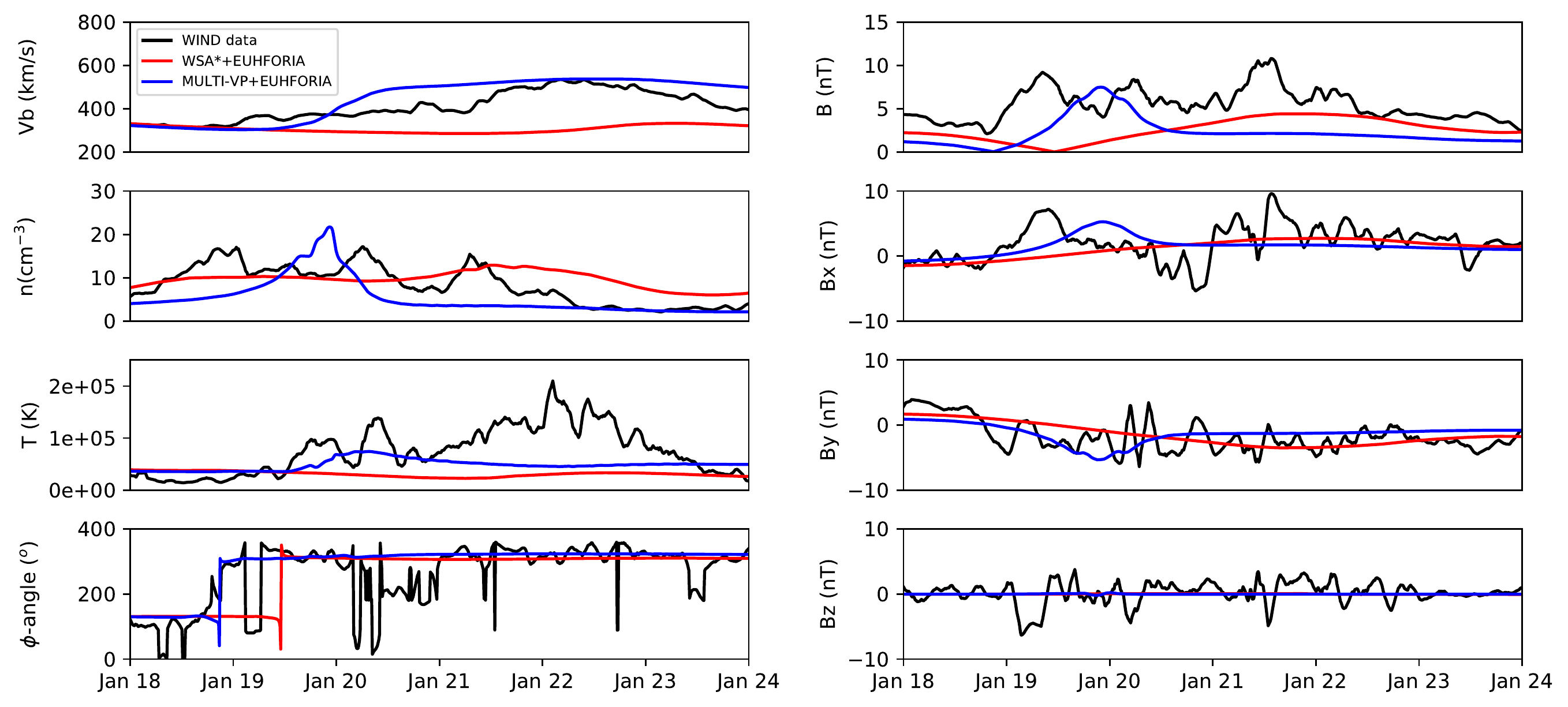}
     \end{subfigure}
 \caption{Plasma and magnetic parameters at $1$~AU as modeled by WSA$^*$+EUHFORIA-heliosphere (red) and MULTI-VP+EUHFORIA-heliosphere (blue) with a GONG magnetogram. The observed data as captured by WIND are depicted in black for the HSS that reached Earth on 2018-01-21.}
\label{Fig:1AU_SolarMin_GONG}
\end{figure*} 


\begin{figure*}[!htb]
\centering
\begin{subfigure}[]{0.48\linewidth}
\includegraphics[width=\linewidth]{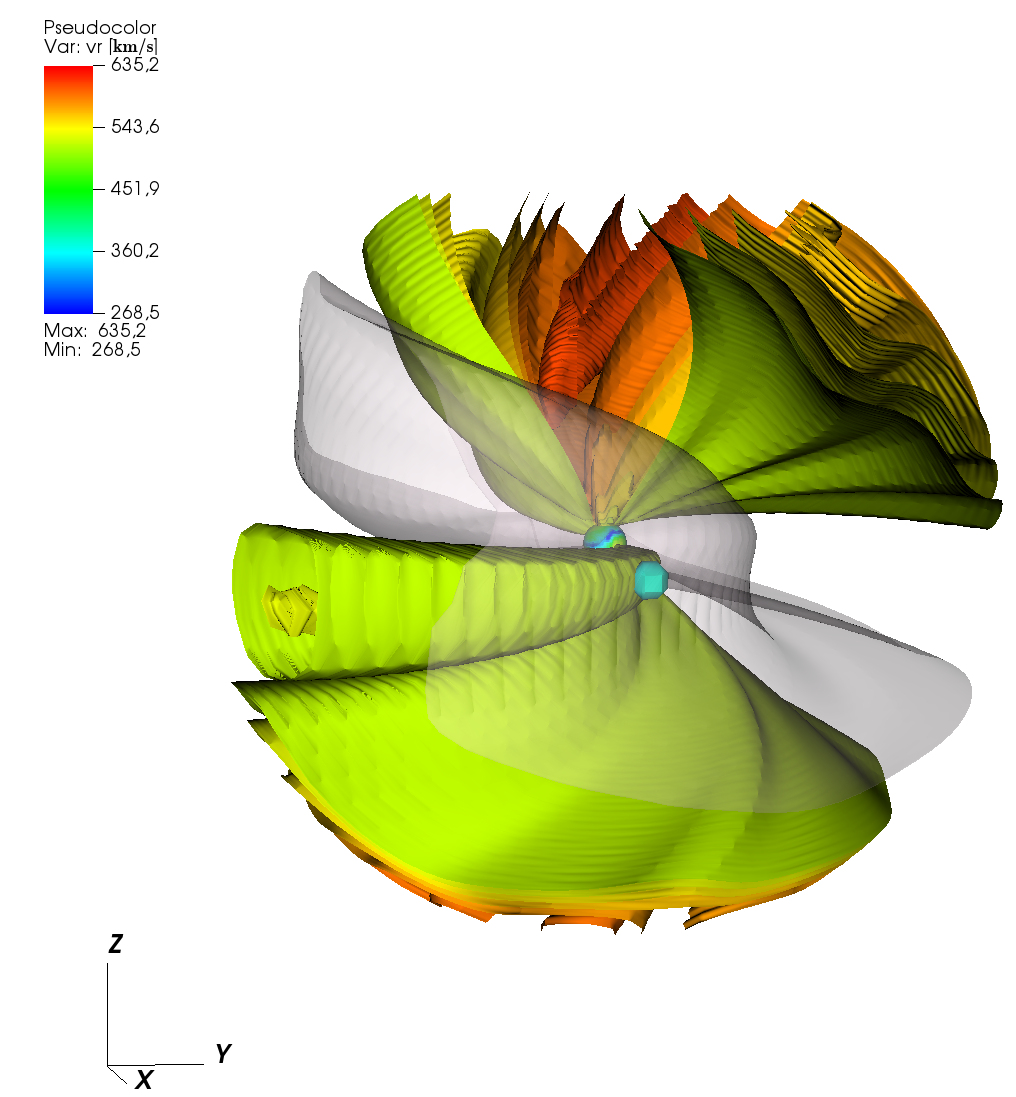}
\caption{MULTI-VP+EUHFORIA}
\end{subfigure}
\hspace{0.01em}%
\begin{subfigure}[]{0.48\linewidth}
\includegraphics[width=\linewidth]{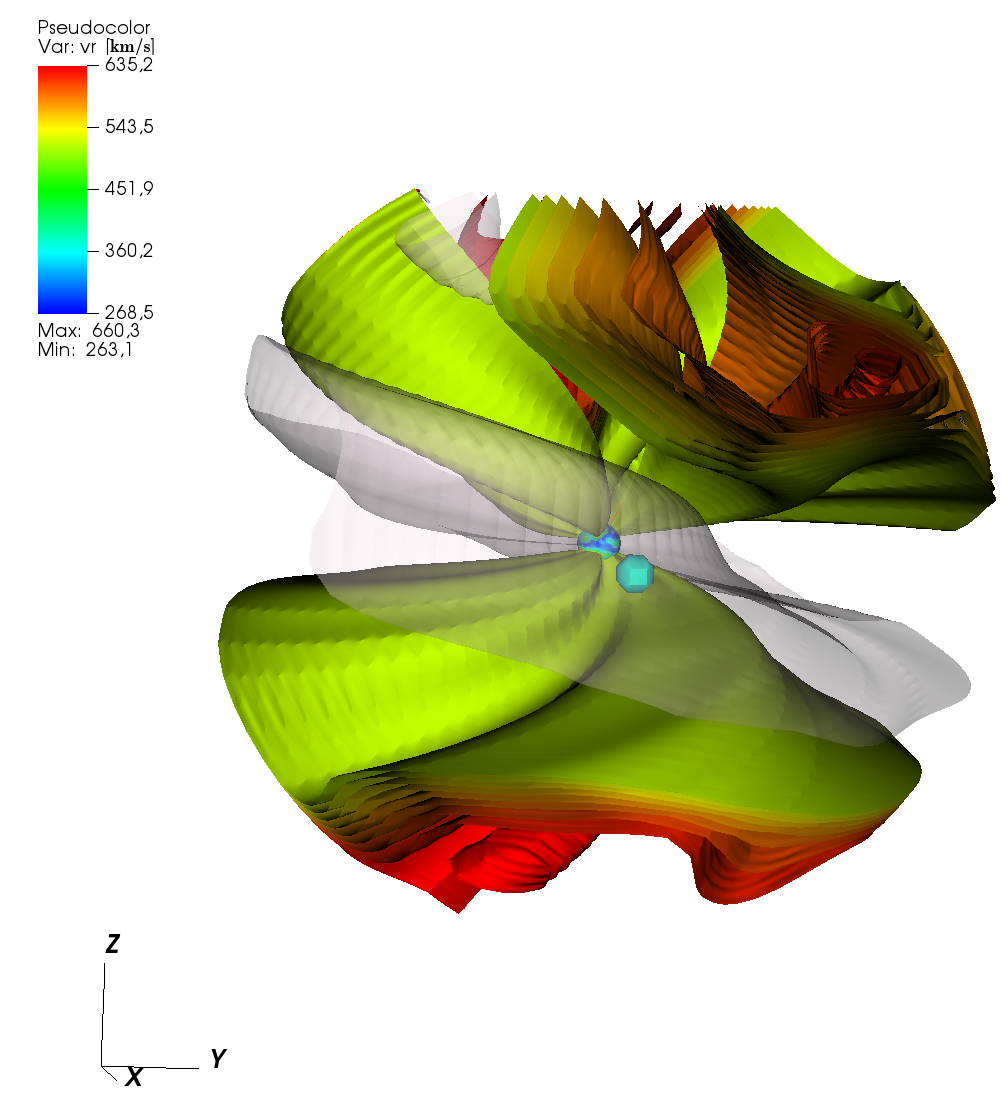}
\caption{WSA$^*$+EUHFORIA}
\end{subfigure}%
\caption{Contour plots of the radial solar wind velocities in 3D space as modeled with a GONG magnetogram (date: 2018-01-17T23:14, CR2199). The range of the velocities shown in the figure is between $[520,600]\;$km/s. The HCS ($B$=0) is depicted in grey while the light-blue sphere represents Earth. The sphere in the center of the figure represents the inner boundary ($0.1$~AU) and is color-coded based on the radial solar wind velocities at that radius, which are provided by the correspondent coronal model each time.}
\label{Fig:VISIT_SolarMin_GONG}
\end{figure*}

\begin{figure*}[!ht]
     \centering
     \begin{subfigure}[b]{0.95\textwidth}
         \centering
         \includegraphics[width=\textwidth]{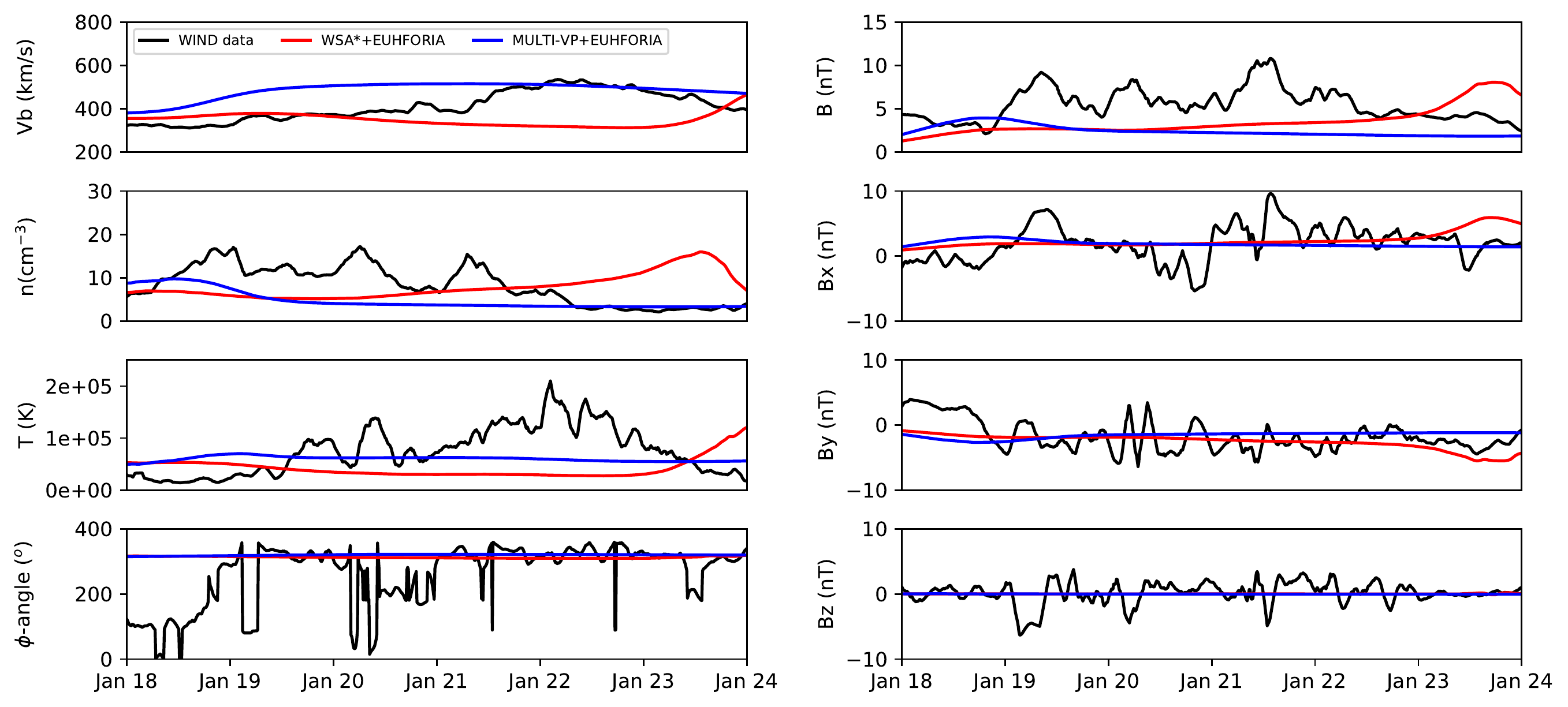}
     \end{subfigure}
 \caption{Plasma and magnetic parameters at 1 AU as modeled by WSA$^*$+EUHFORIA-heliosphere (red) and MULTI-VP+EUHFORIA-heliosphere (blue) with a WSO magnetogram (CR2199). The observed data as captured by WIND are depicted in black for the HSS that reached Earth on 2018-01-21.}
\label{Fig:1AU_SolarMin_WSO}
\end{figure*} 


\begin{figure*}[!htb]
\centering
\begin{subfigure}[]{0.48\linewidth}
\includegraphics[width=\linewidth]{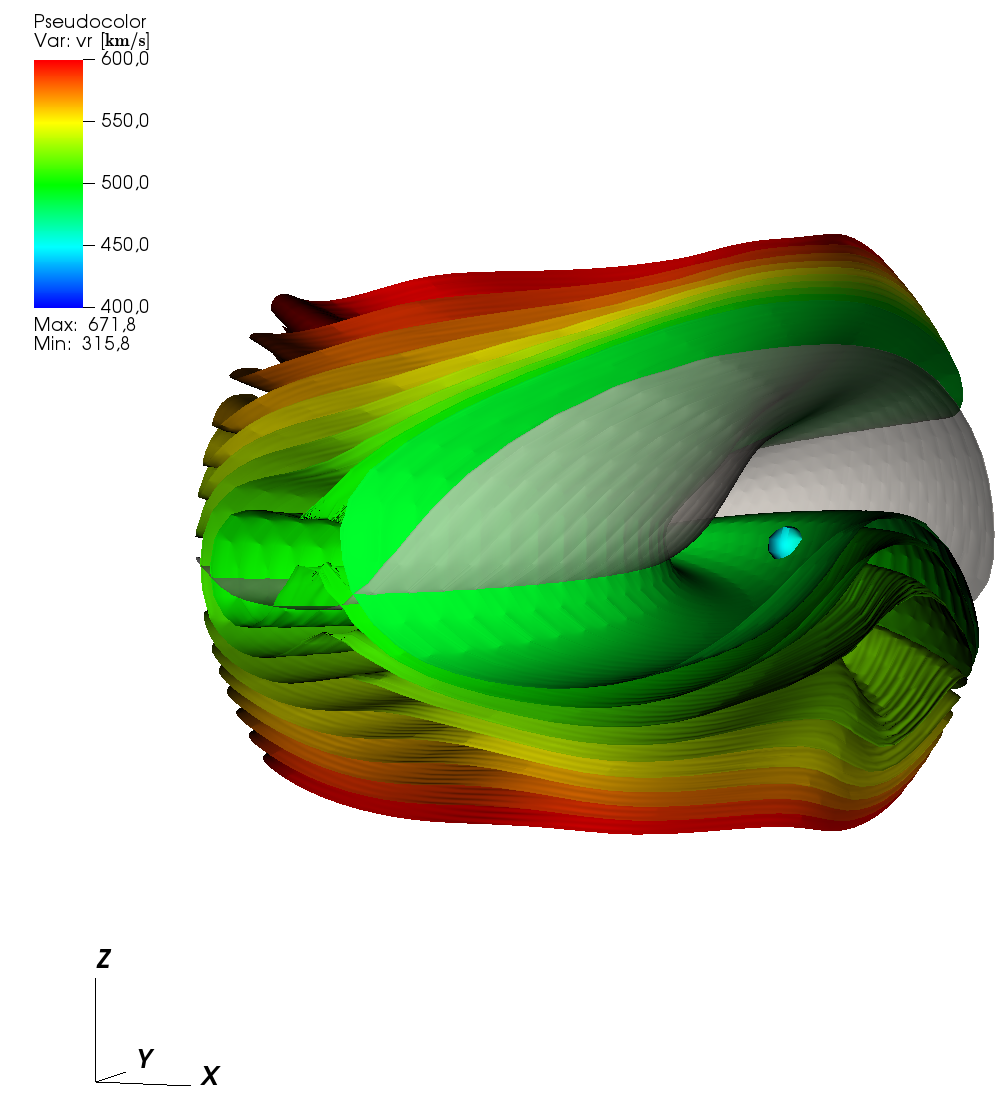}
\caption{MULTI-VP+EUHFORIA}
\end{subfigure}
\hspace{0.01em}%
\begin{subfigure}[]{0.48\linewidth}
\includegraphics[width=\linewidth]{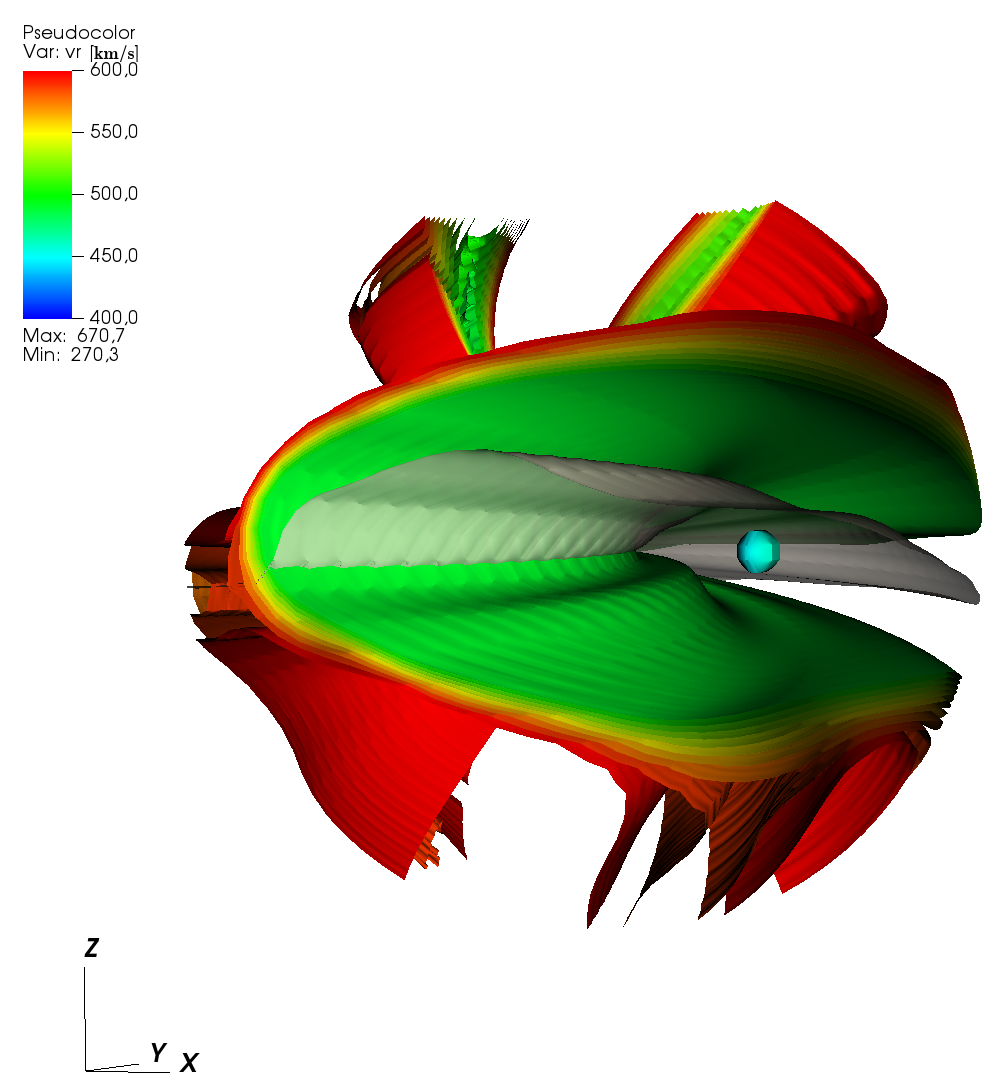}
\caption{WSA$^*$+EUHFORIA}
\end{subfigure}%
\caption{Contour plots of the radial solar wind velocities in 3D space as modeled with a WSO magnetogram (CR2199). The range of the velocities shown in the figure is between $[500,600]\;$km/s. The HCS ($B$=0) is depicted in grey while the light-blue sphere represents Earth. The sphere in the center of the figure represents the inner boundary ($0.1$~AU) and is color-coded based on the radial solar wind velocities at that radius which are provided by the correspondent coronal model each time.}
\label{Fig:VISIT_SolarMin_WSO}
\end{figure*}

\begin{figure*}[ht]
     \centering
     \begin{subfigure}[b]{0.95\textwidth}
         \centering
         \includegraphics[width=\textwidth]{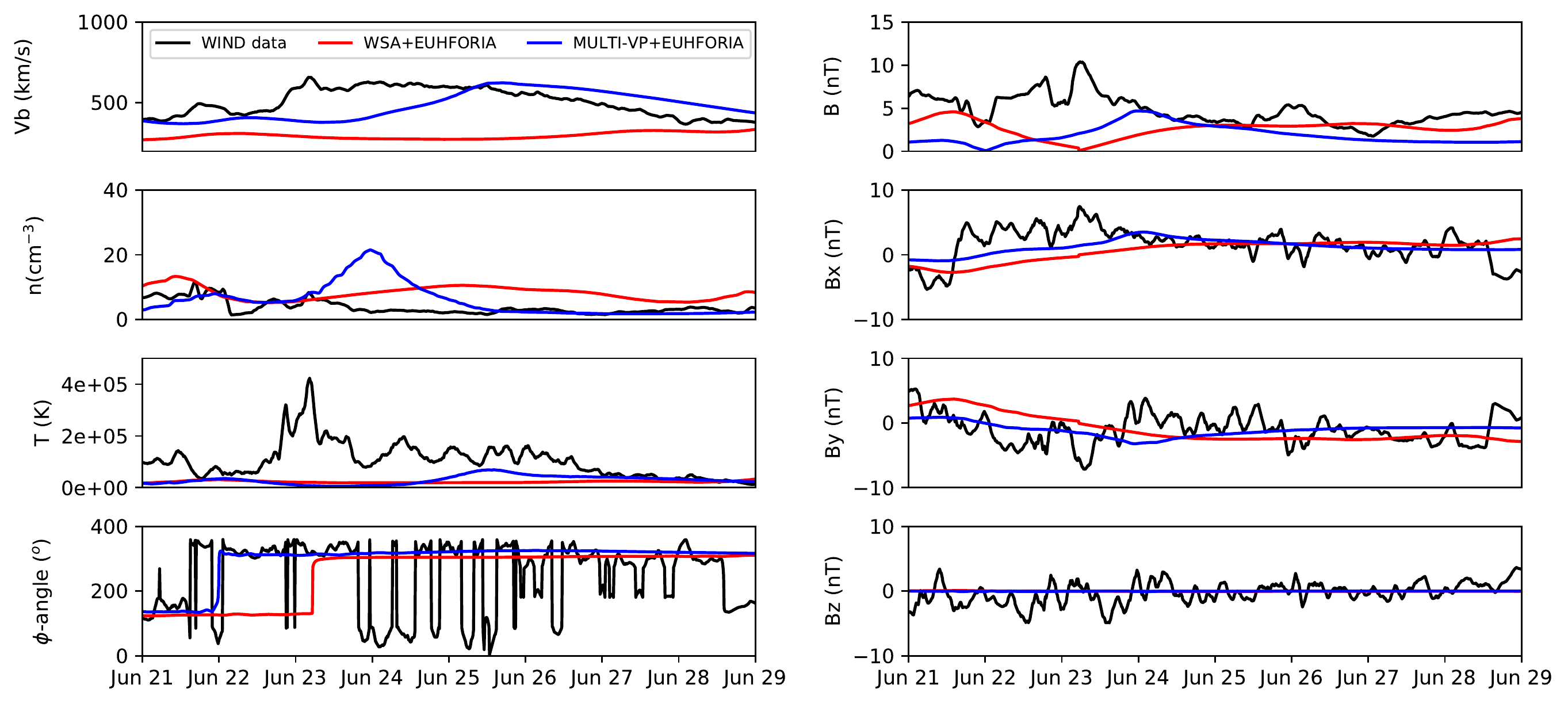}
     \end{subfigure}
 \caption{Plasma and magnetic parameters at 1 AU as modeled by WSA$^*$+EUHFORIA-heliosphere (red) and MULTI-VP+EUHFORIA-heliosphere (blue) with a GONG magnetogram. The observed data as captured by WIND are depicted in black for the HSS that reached Earth on 2011-06-22.}
\label{Fig:1AU_SolarMax_GONG}
\end{figure*} 


\begin{figure*}[!htb]
\centering
\begin{subfigure}[]{0.48\linewidth}
\includegraphics[width=\linewidth]{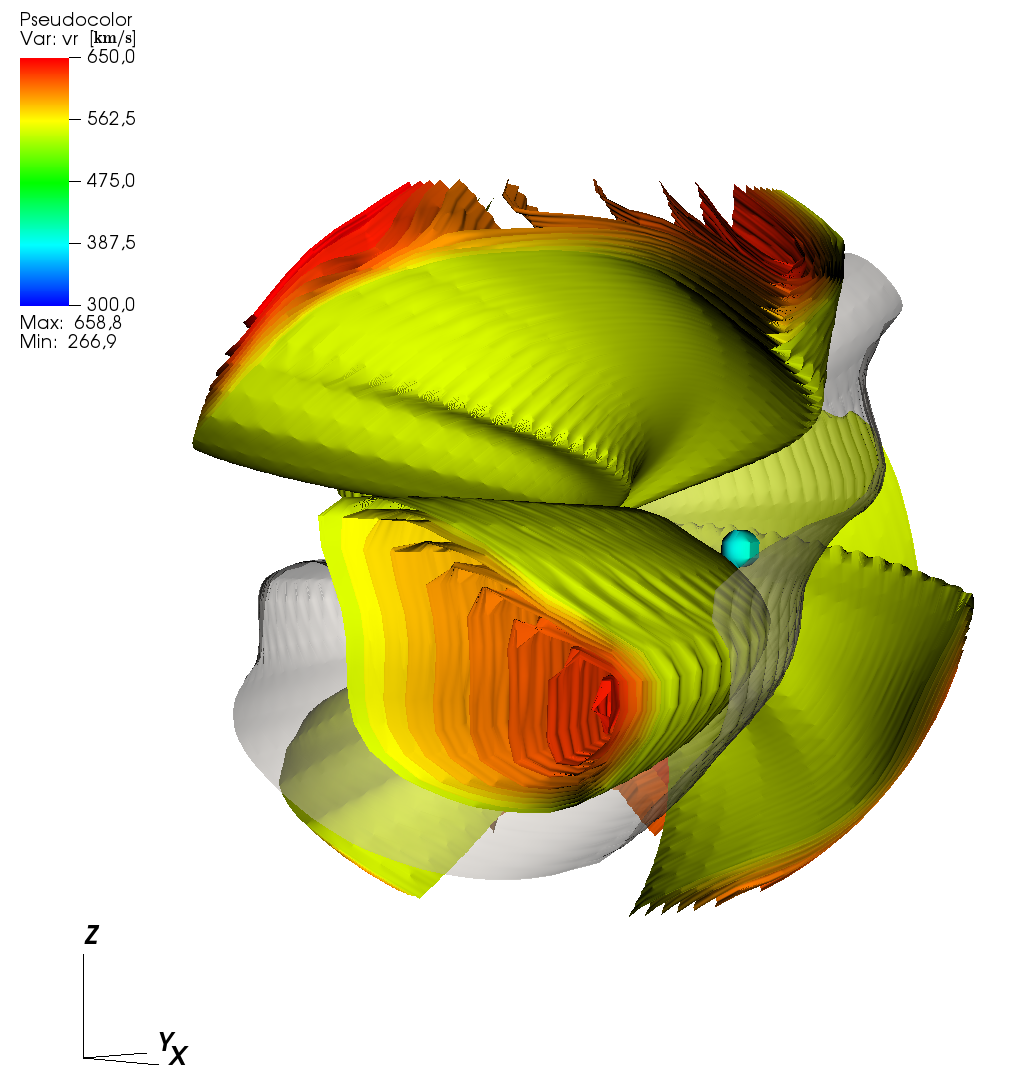}
\caption{MULTI-VP+EUHFORIA}
\end{subfigure}
\hspace{0.01em}%
\begin{subfigure}[]{0.48\linewidth}
\includegraphics[width=\linewidth]{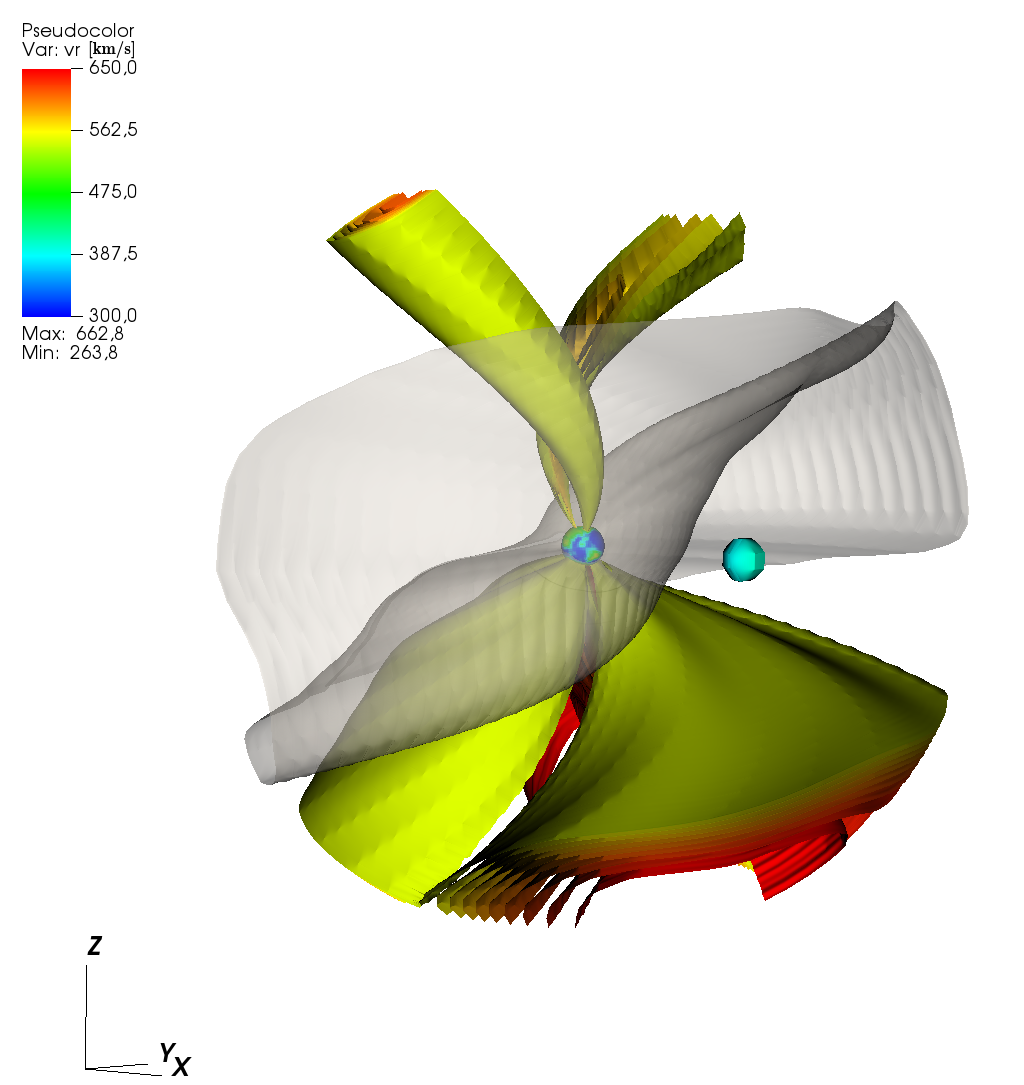}
\caption{WSA$^*$+EUHFORIA}
\end{subfigure}%
\caption{Contour plots of the radial solar wind velocities in 3D space as modeled with a GONG magnetogram (date: 2011-06-20T23:54, CR2111). The range of the velocities shown in the figure is between $[550,650]\;$km/s. The HCS ($B$=0) is depicted in grey while the light-blue sphere represents Earth. The sphere in the center of the figure represents the inner boundary ($0.1$~AU) and is color-coded based on the radial solar wind velocities at that radius which are provided by the correspondent coronal model each time.}
\label{Fig:VISIT_SolarMax_GONG}
\end{figure*}

\begin{figure*}[ht]
     \centering
     \begin{subfigure}[b]{0.95\textwidth}
         \centering
         \includegraphics[width=\textwidth]{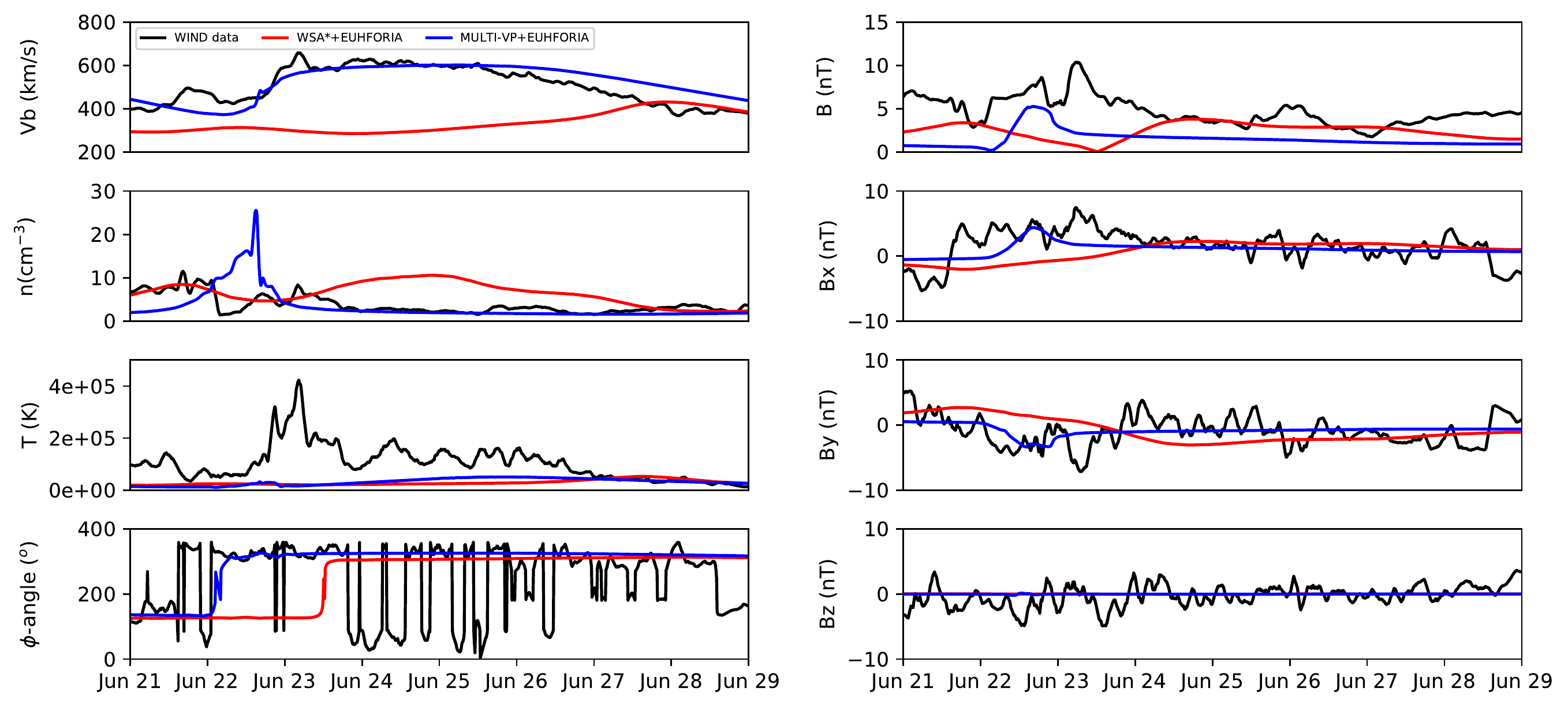}
     \end{subfigure}
 \caption{Plasma and magnetic parameters at 1 AU as modeled by WSA$^*$+EUHFORIA-heliosphere (red) and MULTI-VP+EUHFORIA-heliosphere (blue) with a WSO magnetogram. The observed data as captured by WIND are depicted in black for the HSS that reached Earth on 2011-06-22.}
\label{Fig:1AU_SolarMax_WSO}
\end{figure*} 


\begin{figure*}[!htb]
\centering
\begin{subfigure}[]{0.48\linewidth}
\includegraphics[width=\linewidth]{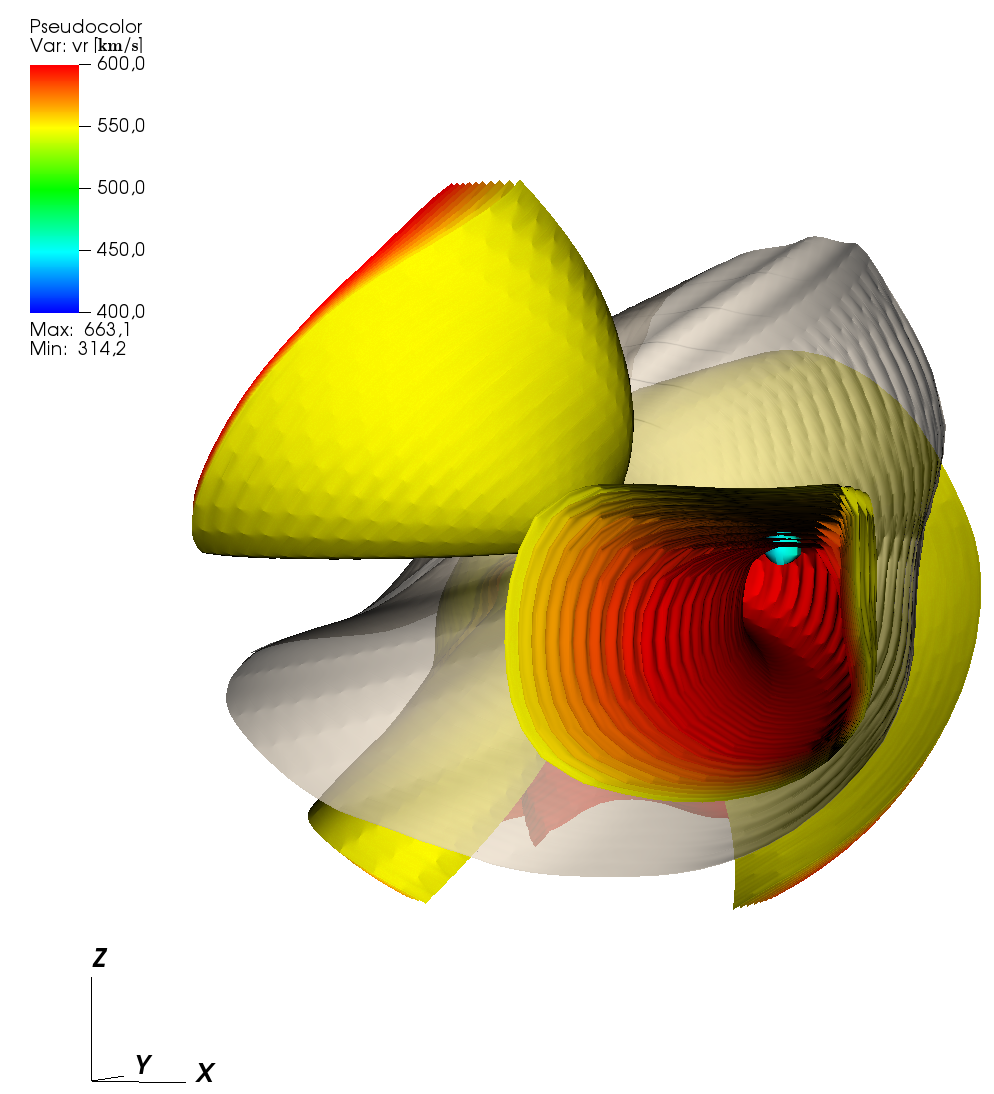}
\caption{MULTI-VP+EUHFORIA}
\end{subfigure}
\hspace{0.01em}%
\begin{subfigure}[]{0.48\linewidth}
\includegraphics[width=\linewidth]{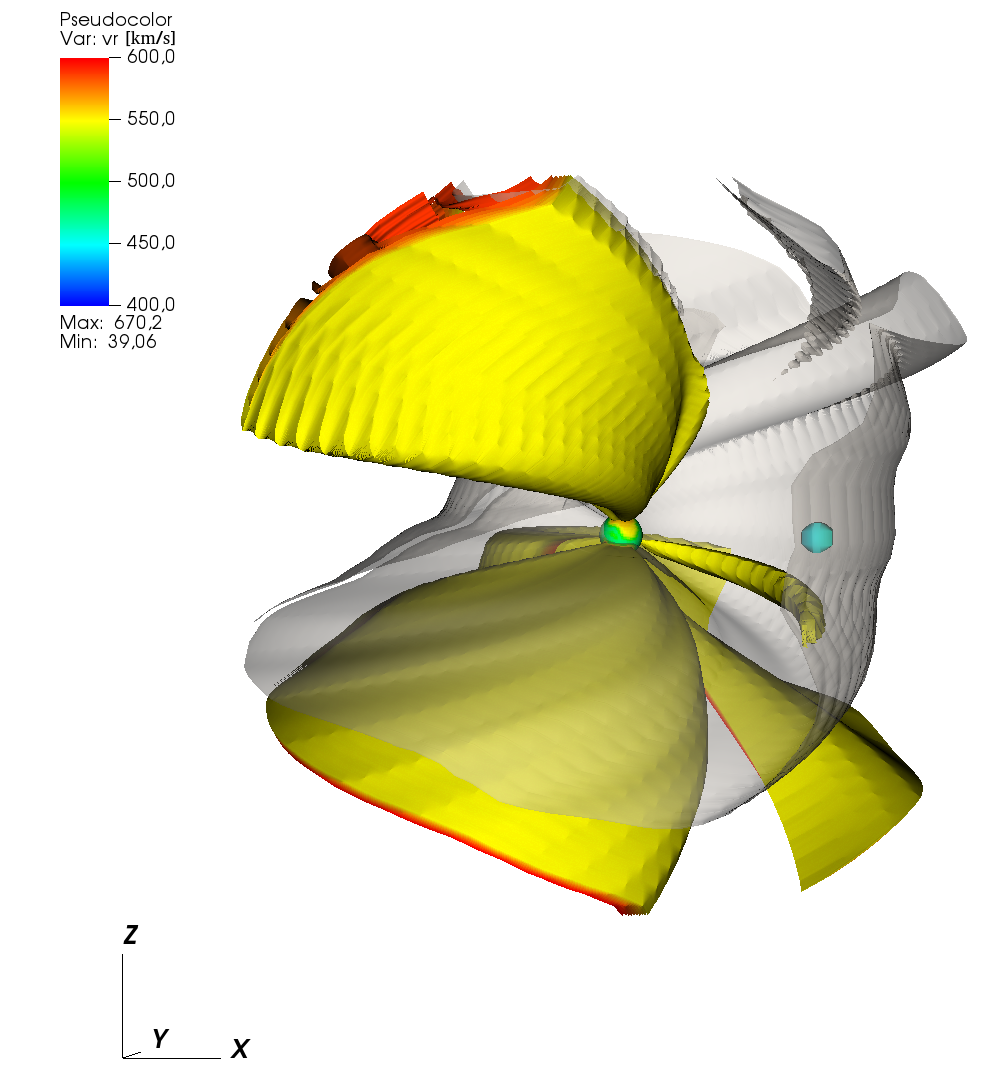}
\caption{WSA$^*$+EUHFORIA}
\end{subfigure}%
\caption{Contour plots of the radial solar wind velocities in 3D space as modeled with a WSO magnetogram (CR2111). The range of the velocities shown in the figure is between $[550,600]\;$km/s. The HCS ($B$=0) is depicted in grey while the light-blue sphere represents Earth. The sphere in the center of the figure represents the inner boundary ($0.1$~AU) and is color-coded based on the radial solar wind velocities at that radius which are provided by the correspondent coronal model each time.}
\label{Fig:VISIT_SolarMax_WSO}
\end{figure*}


\section{MULTI-VP+EUHFORIA-heliosphere versus WSA$^*$+EUHFORIA-heliosphere results at $1$~AU}

In Figs.~\ref{Fig:1AU_SolarMin_GONG}, \ref{Fig:1AU_SolarMin_WSO}, \ref{Fig:1AU_SolarMax_GONG}, and \ref{Fig:1AU_SolarMax_WSO}, the plasma and magnetic properties (bulk speed: $V_{b}$, proton density: $n$, temperature: $T$, interplanetary magnetic field (IMF) $\phi$-angle, magnitude of the IMF: $B$, the three components of the IMF: $B_{x}$, $B_{y}$, $B_{z}$) of the studied HSSs are presented as a function of time in GSE coordinates. The simulations were conducted by assuming a uniform mesh of $4^{\circ}$ in longitude (90 cells), latitude (30 cells) and a radial resolution of 512 cells (radial step of $\Delta r \approx0.0037\;$AU). The boundary values at 0.1 AU are provided at the standard 2$^{\circ}$ x 2$^{\circ}$ resolution by WSA$^*$, no matter the employed magnetogram. In the MULTI-VP case, the boundary values are provided at 2$^{\circ}$ x 2$^{\circ}$ resolution when employing a GONG magnetogram and at 5$^{\circ}$ x 5$^{\circ}$ when using WSO magnetograms, as described in Section 3. Figures~\ref{Fig:VISIT_SolarMin_GONG}, \ref{Fig:VISIT_SolarMin_WSO}, \ref{Fig:VISIT_SolarMax_GONG}, and \ref{Fig:VISIT_SolarMax_WSO} show the simulated solar wind radial velocity in 3D space, presented around the moment when the studied HSS reached Earth.  

\subsection{HSS case during the period of low solar activity}

In Fig.~\ref{Fig:1AU_SolarMin_GONG}, we observe that both the bulk solar wind speed as well as the proton density signatures at Earth are reproduced by MULTI-VP+EUHFORIA-heliosphere (blue line) while this is not the case for the WSA$^*$+EUHFORIA-heliosphere. 
Moreover, results with the former set-up show an increase in temperature above the slow solar wind levels by the time the HSS reaches Earth. Although this increase does not reach the values observed at L1 by the WIND spacecraft, it is closer to the observations than the values obtained by WSA$^*$+EUHFORIA-heliosphere. The IMF $\phi$-angle is captured well by both models, but the polarity of MULTI-VP+EUHFORIA-heliosphere changes earlier. Therefore, it is in better agreement with the observed polarity change on late Jan. 18. The fluctuations of the total magnetic field ($B$) as well as its $B_{x}$ and $B_{y}$ components at the stream interaction, are better reproduced by MULTI-VP+EUHFORIA-heliosphere. However, the models are not expected to reproduce the fluctuations of the IMF $B_{z}$ component since the meridional component of the magnetic field is always set to zero at 0.1 AU. This causes the models to severely underestimate $B_{z}$ throughout the simulation domain. In Fig.~\ref{Fig:VISIT_SolarMin_GONG}, we visualize the simulated radial velocity in 3D space. More specifically, Fig.~\ref{Fig:VISIT_SolarMin_GONG}a shows a HSS arriving at Earth with velocities between $[520, 600]\;$km/s, while in Fig.~\ref{Fig:VISIT_SolarMin_GONG}b, no HSS seems to impact the planet during the period of interest. 

Figure~\ref{Fig:1AU_SolarMin_WSO} shows that neither of the studied models clearly reproduce the HSS, when using a WSO magnetogram (see also Fig.~\ref{Fig:VISIT_SolarMin_WSO}). MULTI-VP overestimates the bulk speed up to at least 100 km/s before the true initiation of the HSS while the $n$ and $T$ signatures do not clearly indicate the HSS arrival. In the WSA$^{*}$ case, the modeled solar wind parameters are underestimated and do not follow any of the observed HSS patterns.

\subsection{HSS case during the period of high solar activity}

Figure~\ref{Fig:1AU_SolarMax_GONG} shows the comparison of the simulated and observed solar wind parameters for the HSS during the period of high solar activity (reached Earth on 2011-06-22) with a GONG magnetogram. We notice that the HSS, as simulated by MULTI-VP+EUHFORIA-heliosphere, arrives approximately two days later compared to WIND observations. The amplitude of the modeled velocity reaches the observed values, but the peak in density is overestimated. On the other hand, very small increase in the modeled values of temperature and the total magnetic field is detected compared to WIND measurements. The $B_{x}$ and $B_{y}$ components do not comply with observations while $B_{z}$ is not expected to be reproduced by the model, as already mentioned in the previous subsection.

WSA$^*$+EUHFORIA-heliosphere did not reproduce the increase in the solar wind speed and temperature. The density is closer to the observed values between 21-22 June but it gets overestimated (by $\approx$ 5-10 cm$^{-3}$) after that date. The IMF-$\phi$-angle has the correct polarity but it arrives $\approx$1 day later than in MULTI-VP+EUHFORIA-heliosphere case. Also, the $B_{x}$ and $B_{y}$ components do not differ much from the ones reproduced by the newly coupled set-up. Figure~\ref{Fig:VISIT_SolarMax_GONG} shows the modeled radial solar wind speed in 3D space. This presentation clearly shows the difference in the simulation results presented also in the $vr$ plot of Fig.~\ref{Fig:1AU_SolarMax_GONG}. The WSA$^*$+EUHFORIA-heliosphere output does not show any HSS arriving at Earth while MULTI-VP+EUHFORIA-heliosphere reproduced the HSS with velocities between $[550,650]\;$km/s directly impacting the planet. 

In Fig.~\ref{Fig:1AU_SolarMax_WSO} and \ref{Fig:VISIT_SolarMax_WSO}, the same quantities are compared for the HSS observed during the high levels of solar activity, using a WSO magnetogram. Figure~\ref{Fig:1AU_SolarMax_WSO} shows that MULTI-VP+EUHFORIA-heliosphere accurately captures the arrival time and velocity amplitude of the HSS, opposite to WSA$^*$+EUHFORIA-heliosphere. After plotting radial velocities between $[550,600]\;$km/s in 3D space as shown in Fig. \ref{Fig:VISIT_SolarMax_WSO}b, we only identify a stream surpassing Earth from the southern part, which does not affect the planet. On the other hand, the coupled MULTI-VP+EUHFORIA-heliosphere model yields an extended HSS directly impacting Earth, which is in accordance with the blue time series observed in the first panel of Fig.~\ref{Fig:1AU_SolarMax_WSO}. Moreover, MULTI-VP+EUHFORIA-heliosphere overestimated the expected peak in the proton density, while WSA$^*$+EUHFORIA-heliosphere diverges from the observations, especially after the arrival of the HSS at Earth. Neither of the models reproduced the large temperature increase during the advent of the HSS at Earth. The modeled IMF $\phi$-angle change occurred late compared to observations for both models, though the simulated polarities before and after the change of the $\phi$-angle were correct. The magnetic field fluctuations (except the $B_{z}$ component) are captured better by the MULTI-VP+EUHFORIA-heliosphere model than by the WSA$^*$+EUHFORIA-heliosphere. It is, however, notable that the magnetic field magnitude is consistently smaller in the simulations as compared to WIND data, for both models.

\begin{figure*}[ht]
     \centering
     \begin{subfigure}[b]{0.75\textwidth}
         \centering
         \includegraphics[width=\textwidth]{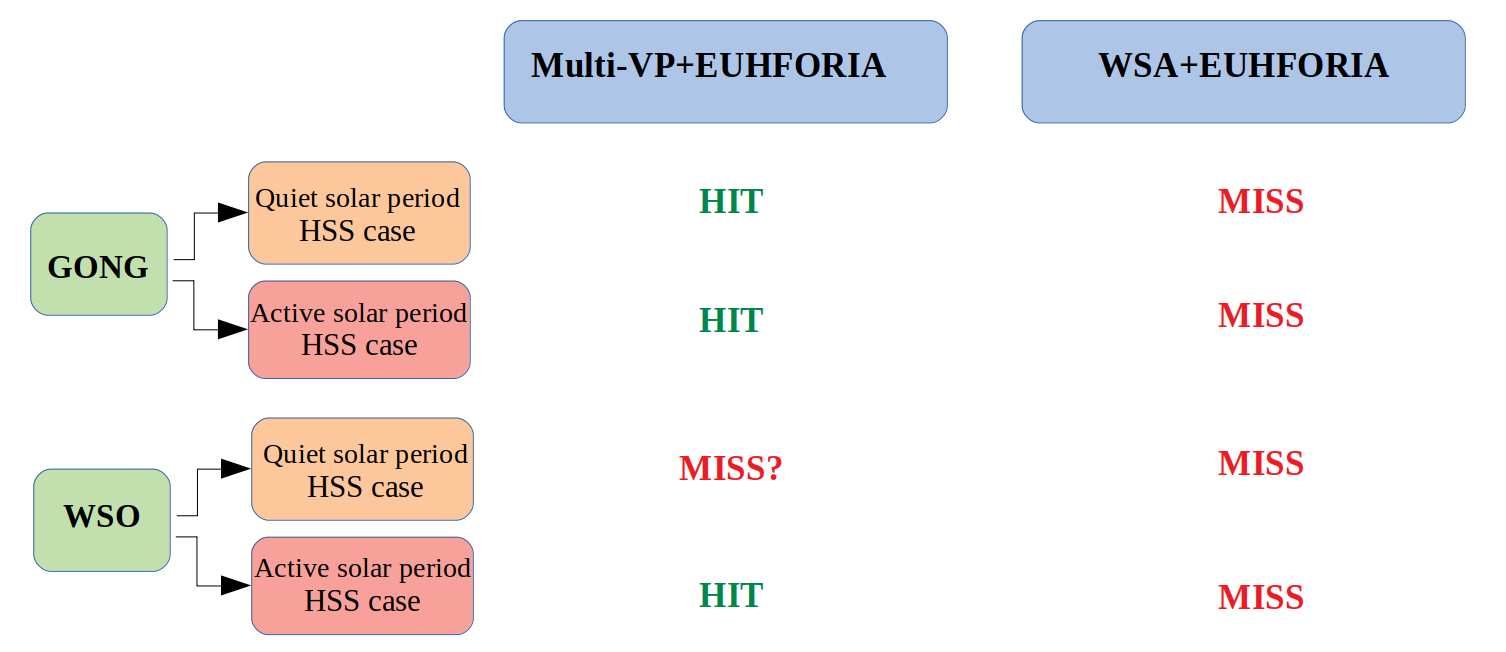}
     \end{subfigure}
 \caption{Summary of the modeling results at $1$~AU regarding the solar wind bulk speed.}
\label{Fig:HitMissesChart}
\end{figure*} 

\section{Summary and Discussion}
In this study, we implemented for the first time an alternative coronal model in EUHFORIA, the MULTI-VP model \citep{Pinto17}. We compared the output of the default coronal model with the output from MULTI-VP at the inner boundary of the heliospheric domain of EUHFORIA in order to understand differences between the two models, before they propagate to Earth. We also compared the performance of WSA$^{*}$+EUHFORIA-heliosphere and MULTI-VP+EUHFORIA-heliosphere against in situ observations at Earth. In the frame of this study, we considered two different HSS cases, one during a period of low solar activity and one during a period of high solar activity. We also employed two different magnetograms, i.e., GONG and WSO. Our results show that the choice of both the coronal model and the magnetogram play an important role on the accuracy of the solar wind prediction. However, it is not clear which component plays the most important role for the modeled results obtained at Earth. A statistical analysis with an appropriate number of simulations is needed to confirm our findings. 

In the process of implementing MULTI-VP model in EUHFORIA, we encountered a number of elemental flows that are sub-critical at $0.1$~AU (typically less than $1\%$ of the whole map, and up to a few percent in the most extreme cases). MULTI-VP cannot assure \emph{a priori} that the solar wind solutions it computes are super-fast at all angular positions at the target altitude of $0.1$~AU, as required for EUHFORIA. To correctly feed the MULTI-VP data into the heliospheric part of EUHFORIA, we needed to transform these speeds to super-critical such that all MHD characteristic curves are out-going at 0.1 AU \citep{StefaansBook}. The correction was done by interpolating the sub-critical fast magnetosonic pixels with their closest super-critical neighbors obeying the mass-flux conservation. Once super-criticality has been achieved at the boundary, we were able to study the $v_r$, $n$, $T$, $B_{r}$ maps there. The analysis of inner boundary maps allowed us to obtain a first-order estimation regarding the differences between the models. Moreover, it helped us understand how the two coronal models deal with different magnetograms.

In Fig.~\ref{Fig:HitMissesChart}, we outline our conclusions based on the modeled bulk speed signatures at $1$~AU. The results show that MULTI-VP+EUHFORIA-heliosphere is able to reproduce both HSSs cases when using GONG magnetograms as well as the HSS case during the active solar period, when employing a WSO magnetogram. It is not certain, though, if it captures the HSS during the period of low solar activity when using the latter type of magnetogram. This is the reason we describe it as a "miss" with a question mark in Fig.~\ref{Fig:HitMissesChart}. The WSA$^{*}$+EUHFORIA-heliosphere combination, which we use as the reference model, does not reproduce any of the two test-case HSSs, regardless the magnetogram. However, these HSSs were specifically chosen in purpose as cases that we knew a priori that were not reproduced well by the default EUHFORIA set-up, in order to test the performance of MULTI-VP in combination with EUHFORIA-heliosphere. A bigger sample of HSSs needs to be simulated in order to determine if one of the models consistently outperforms the other.

The main reason that the two models provide different results at $1$~AU, given the same input magnetogram, is the way they calculate the solar wind state at 0.1 AU. Even though both coronal models use the PFSS model to reconstruct the magnetic field in the low corona, they rely on different techniques to reconstruct the magnetic field higher in the corona and up to the radial distance of $0.1$~AU. The default EUHFORIA set-up is based on the SCS model to create a more uniform magnetic field, and on the WSA speed (Eq.~\ref{WSA_vr}), which determines the solar wind plasma and magnetic parameters at $0.1$~AU. The WSA wind speed essentially depends on magnetic information at two specific altitudes, the solar surface and $0.1$~AU (see Eqs.~\ref{fte_factor} and \ref{WSA_vr}). On the other hand, MULTI-VP calculates the heating and acceleration of all wind streams at every height based on Eqs.~6, 7, 8 and provides a uniform magnetic field away from the Sun by applying an additional flux-tube expansion profile to them. Therefore, the differences in the numerical approach and underlying assumptions of the two models lead to distinctly different output. It is also important to mention that even though WSA$^{*}$ does not reproduce the two particular HSSs in this study, it is considered a reliable coronal model that is computationally inexpensive, in comparison to MULTI-VP.

\begin{acknowledgements}
    \textbf
    The authors acknowledge the anonymous referee for the constructive comments that helped improving the manuscript. They also extend their acknowledgements to Jens Pomoell from the University of Helsinki for fruitful discussions on this work. E.S. and I. C.J. were supported by PhD grants awarded by the Royal Observatory of Belgium. C.S. acknowledges funding from the Research Foundation - Flanders (FWO, fellowship no. 1S42817N). N.W. acknowledges funding from the Research Foundation - Flanders (FWO, fellowship no. 1184319N). S.P. was also supported by the projects C14/19/089 (C1 project Internal Funds KU Leuven), G.0D07.19N (FWO-Vlaanderen), SIDC Data Exploitation (ESA Prodex-12). EUHFORIA is developed as a joint effort between the KU Leuven and the University of Helsinki. The validation of solar wind and CME modeling with EUHFORIA is being performed within the BRAIN-be project CCSOM (Constraining CMEs and Shocks by Observations and modeling throughout the inner heliosphere; www.sidc.be/ccsom/) and BRAIN-be project SWiM (Solar Wind Modeling with EUHFORIA for the new heliospheric missions. This project has received funding from the European Union’s Horizon 2020 research and innovation programs under grant agreements No 870405 (EUHFORIA 2.0) and 870437 (SafeSpace). Computational resources and services used in this work were provided by the VSC (Flemish Supercomputer Center), funded by the Research Foundation Flanders (FWO) and the Flemish Government-Department EWI. The MULTI-VP numerical simulations were performed using HPC resources from CALMIP (Grant 2020-P1504).
     
\end{acknowledgements}

\bibliographystyle{aa}
\bibliography{bibliography}

\end{document}